\documentclass[letter]{aa}
\usepackage[varg]{txfonts}

\usepackage{hyperref}
\hypersetup{colorlinks=true,linkcolor=blue,citecolor=blue,filecolor=blue,urlcolor=blue}

\usepackage{tikz}
\usepackage{graphicx}
\graphicspath{{/nbody/betelgeuse/},{images/}}

\usepackage{bm}

\usepackage{marvosym}
\definecolor{Mercury}{HTML}{ABABAB}
\definecolor{Venus}{HTML}{0B610B}
\definecolor{Earth}{HTML}{0000FF}
\definecolor{Mars}{HTML}{FF0000}
\definecolor{Jupiter}{HTML}{FF8000}
\definecolor{Saturn}{HTML}{61380B}
\definecolor{Uranus}{HTML}{00BFFF}
\definecolor{Neptune}{HTML}{EE7FEE}

\newcommand{\varEarth}{{\small\boldmath$\oplus$\unboldmath}}

\newcommand{\pc}{\mathrm{pc}}
\newcommand{\au}{\mathrm{au}}
\newcommand{\yr}{\mathrm{yr}}
\newcommand{\Myr}{\mathrm{Myr}}
\newcommand{\Msun}{M_\odot}
\newcommand{\kms}{\mathrm{km\,s^{-1}}}

\newcommand{\adeg}{^\circ}

\newcommand{\aOri}{$\alpha$\,Ori}
\newcommand{\I}{\mathrm{I}} 
\newcommand{\BH}{\mathrm{BH}}
\newcommand{\NS}{\mathrm{NS}}
\newcommand{\disp}{\sigma_\mathrm{k}}
\newcommand{\inc}{\theta}
\newcommand{\retain}{\mathrm{ret}}
\newcommand{\capture}{\mathrm{cap}}
\newcommand{\grav}{\mathrm{grav}}
\newcommand{\Np}{N_\mathrm{p}}

\renewcommand{\vec}[1]{\bm{#1}}

\definecolor{lime}{HTML}{A6CE39}
\DeclareRobustCommand{\orcidicon}{%
        \begin{tikzpicture}
        \draw[lime, fill=lime] (0,0) 
        circle [radius=0.16] 
        node[white] {{\fontfamily{qag}\selectfont \tiny ID}};
        \draw[white, fill=white] (-0.0625,0.095) 
        circle [radius=0.007];
        \end{tikzpicture}
        \hspace{-2mm}
}

\newcommand{\orcidVP}{\href{https://orcid.org/0000-0002-3031-062X}{\orcidicon}}
\newcommand{\orcidSS}{\href{https://orcid.org/0000-0003-1677-8004}{\orcidicon}}

\begin{document}

\title{Close encounters with the Death Star: Interactions between collapsed bodies and the Solar System}
\subtitle{}

\titlerunning{Encounters between collapsed bodies and the Solar system}

\setcounter{footnote}{4}

\author{V\'aclav Pavl\'ik \inst{\ref{iu},\ref{asu},}\thanks{\email{vpavlik@iu.edu}} \orcidVP
\and Steven N. Shore \inst{\ref{pisa},\ref{infn}} \orcidSS}

\institute{Indiana University, Department of Astronomy, Swain Hall West, 727 E 3$^\text{rd}$ Street, Bloomington, IN 47405, USA \label{iu}
\and Astronomical Institute of the Czech Academy of Sciences, Bo\v{c}n\'i~II~1401, 141~31~Prague~4, Czech Republic \label{asu}
\and Dipartimento di Fisica, Universit\`a di Pisa, largo B. Pontecorvo 3, Pisa 6127, Italy \label{pisa}
\and INFN, Sezione di Pisa, largo B. Pontecorvo 3, Pisa 6127, Italy \label{infn}}

\authorrunning{Pavl\'ik \& Shore}

\date{Received: January 30, 2021 / Accepted: March 22, 2021}

\abstract
{}
{We aim to investigate the consequences of a~fast massive stellar remnant -- a~black hole (BH) or a~neutron star (NS) -- encountering a~planetary system.}
{We modelled a~close encounter between the actual Solar System (SS) and a~$2\,\Msun$ NS and a~$10\,\Msun$ BH, using a~few-body symplectic integrator. We used a~range of impact parameters, orbital phases at the start of the simulation derived from the current SS orbital parameters, encounter velocities, and incidence angles relative to the plane of the SS.}
{We give the distribution of possible outcomes, such as when the SS remains bound, when it suffers a~partial or complete disruption, and in which cases the intruder is able to capture one or more planets, yielding planetary systems around a~BH or a~NS. We also show examples of the long-term stability of the captured planetary systems.}
{}


\maketitle

\section{Introduction}
\label{sec:intro}


Observations have shown that the compact objects produced by supernova (SN) explosions -- neutron stars (NSs) or black holes (BHs) -- can receive natal recoil impulses due to intrinsic asymmetries of their birth event. The best-fit observational velocity distribution of these kicks for NSs is a~Maxwellian \citep{kicks_ns,bh_ns_kicks} with a~mean velocity about $300\,\kms$ and a~dispersion of about $\disp=190\,\kms$. A~BH could receive a~kick of the same magnitude as a~NS \citep{bh_ns_kicks,Repetto2017}; although, more recent investigations argue that lower velocities, for example ${\sim}50\,\kms$, may be needed to explain the retention fraction of BHs in Galactic globular clusters \citep{peuten,baumgardt_sollima,pavl_bh}, and that they are likely reduced by the fallback of mass onto the BH remnant \citep[e.g.][]{belczynski,fryer}.

In this work, we focus on the dynamical consequences of a~passage of a~collapsed fast moving massive body close to a~planetary system using the current Solar System (SS) for realistic initial conditions. Such an event, which could more likely play out in the environment of a~star cluster or an association, has hitherto not been fully examined.

Published studies of the stability of planetary systems undergoing stellar encounters have analysed isolated in-plane events involving low-mass single field stars or binaries
\citep[e.g.][]{li_adams,malmberg_etal} or have been sited in clusters with the impactor velocity taken from the virialised velocity dispersion using Monte Carlo or $N$-body procedures on fictitious planetary systems \citep[e.g.][]{2017MNRAS.470.4337C,wang_etal}.  Our study differs in assuming (1) the real SS with all eight planets, (2) arbitrary off-plane interactions, (3) a~massive intruder of high mass ratio, and (4) a~broad range of impact parameters and encounter velocities. The only study similar in spirit to ours, by \citet{laughlin_adams}, imposed an approximation of the SS and included high-velocity encounters with a~low-mass binary system as the impactor.

\section{Simulation methods}
\label{sec:methods}

The set of nearly 2\,000 numerical simulations we performed in this study examined two encounter scenarios -- BH and NS.
We varied the impact parameter, $b_\I$ (defined in the plane of the SS), the encounter velocity, $v_\I$, the intruder mass, $m_\I$, and incidence angle, $\inc_\I$ (relative to the invariable plane of the SS); the subscript `I' specifies the nature of the impactor: BH and NS. No relativistic effects were considered, so the designations BH and NS are formal and indicate a~point mass.

Our simulations were performed using \texttt{REBOUND} \citep{rebound}, a~symplectic $N$-body code, with a~$15^\mathrm{th}$ order Gauss--Radau integrator \texttt{IAS15} \citep{reboundias15}, which was implemented for \texttt{Python}, and they were analysed using \texttt{Python} with \texttt{NumPy} \citep{numpy} and \texttt{Matplotlib} \citep{matplotlib}. The system was initiated in isolation, ignoring the Galactic gravitational potential and the very low stellar density in the Solar neighbourhood. The duration of the interaction is much lower than the Galactic orbital timescale of the whole system. The initial conditions for specific encounter scenarios are described below in further detail.

\subsection{Initial conditions}

We adopted the same range of impact parameters in both scenarios, $10^{-1} \leq b_\I/\pc \leq 10^{-6}$ (the smallest, $0.3\,\au$, is approximately the current orbit of Mercury). The SN kicks may not be the only reason for these encounters, they may also arise in a~dynamical system \citep[e.g. a~star cluster or an association, as analysed by numerical models][]{2009ApJ...697..458S,2013MNRAS.433..867H,2015MNRAS.453.2759Z,2016ApJ...816...59S,2017MNRAS.470.4337C}, so we chose velocities 10, 50, and $100\,\kms$ in both scenarios. We also included $360\,\kms$ for the NS to account for possibly large velocity kicks or hyper-velocity stars. We also varied the impactor incidence angle, $\inc_\I$, between $0\adeg$ and $180\adeg$ in $30\adeg$ increments.
In each setup, we launched the impactor from $0.5\,\pc$ towards the Sun and ended the simulation when the separation of these two bodies was again greater than $0.5\,\pc$, since there was no effect at a larger distance.\!\footnote{For two models, we also examined the longer-term stability of the captured systems, see Sect.~\ref{sec:results}.}

The initial positions of the planets and the Sun were taken from NASA JPL HORIZONS system for Jan\,1,\,2000 and then varied in each realisation by letting the SS evolve for a~certain amount of time before the start of the simulation.
We adopted nine pre-evolved states, evenly distributed from 0 to 165 years (approximately the orbital period of Neptune), producing an ensemble of possible encounters of the remnant with large planets.
Such a~coarse time step can initially create degenerate positions of the terrestrial planets but can be justified due to the following reasons:
\begin{enumerate}
	\item We let the massive body travel at different velocities from the same distance, so the positions of planets during the closest approach are different despite starting from the same pre-evolved state.
	\item We made several simulations without terrestrial planets. Using deterministic symplectic methods allowed us to make one-to-one comparisons to the simulations with all planets and the same initial configuration. The results show that large gas planets are the dominant agents when it comes to the stability of the SS or capture of planets. Low-mass terrestrial planets being retained, captured, or ejected is a~secondary outcome that influences the orbital parameters of the final system should they be captured or retained. Hence, the terrestrial planets behave essentially as test masses. We plotted one such comparison in Fig.~\ref{fig:detail_4-8pl} -- the simulations are almost identical except for the capture of Venus and Earth and a~slightly different semi-major axis of Saturn.
	\item We also performed high-resolution calculations of the BH encounter, with a~fixed incidence angle and a~time step between realisations of 0.5 year (i.e. 330 simulations in total, see Appendix~\ref{ap:bet}).
\end{enumerate}

\subsection{Orbital parameters}

For analysing the simulation outputs, we distinguished planets that remained bound to the Sun from those captured by the intruder. We counted all the orbiting bodies within a~radius of 200\,au around the relevant central object that remained bound until the end of the simulation. The orbital parameters are defined as follows. The semi-major axis is given by
\begin{equation}
	\label{eq:semi}
	a = \left( \frac{2}{|\vec{r}|} - \frac{|\vec{v}|^2}{GM} \right)^{-1} \,,
\end{equation}
where $\vec{r}$ and $\vec{v}$ are the radius and velocity vectors, respectively, in the co-moving frame of the central body (the Sun or the impactor), $G$ is the Newtonian gravitational constant, and $M$ is the mass of the central body. The orbital inclination is defined by
\begin{equation}
	\label{eq:inc}
	i = \arccos{\Big( h_z \,\big/\, |\vec{h}| \Big)} \,,
\end{equation}
where $\vec{h} = \vec{r}\times\vec{v}$ in the co-moving frame, and the eccentricity vector is
\begin{equation}
	\label{eq:ecc}
	\vec{e} = \frac{\vec{v}\times\vec{h}}{GM} - \frac{\vec{r}}{|\vec{r}|} \,,
\end{equation}
where the orbital eccentricity is $e = |\vec{e}|$.
These values are plotted for each planet (labelled as Mercury~\Mercury, Venus~\Venus, Earth~\varEarth, Mars~\Mars, Jupiter~\Jupiter, Saturn~\Saturn, Uranus~\Uranus,\ and Neptune~\Neptune) that has been captured or retained in the top six rows of Figs.~\ref{fig:bh} \&~\ref{fig:ns} and in Figs.~\ref{fig:bet}, \ref{fig:bh_semi}--\ref{fig:bh_sun_inc}, \&~\ref{fig:ns_semi}--\ref{fig:ns_sun_inc}. We emphasise that these plots show the distribution of orbital parameters, but not the actual number of captured or retained planets, which is plotted, for example, in Figs.~\ref{fig:bh_cap}, \ref{fig:bh_sun_ret}, \ref{fig:ns_cap}, \&~\ref{fig:ns_sun_ret}.

\section{Results}
\label{sec:results}

Encounters with field stars have been simulated previously, but none specifically assumed that the intruder is a~massive compact object.
We therefore explored its interaction with a~planetary system for arbitrary incidence angles which is likely in denser environments, for example, star clusters \citep{Varri2018} or the Galactic bulge, where thousands of NSs may be present \citep[e.g.][]{msp}.
The encounter velocities originating from SN kicks will be significantly higher than the virialised velocities that have been used in published simulations \citep[e.g.][]{wang_etal}. In fact, they are closer to what would be expected for a~high-velocity star in the Solar neighbourhood \citep{laughlin_adams}.
An estimate of the gravitational capture distance, assuming that the intruder can be bound in the centre of mass frame, is given by $r_{\grav,\I} = 2 G \left( m_\I + \Msun \right) \big/\, v_\I^2$\,. This scales as
\begin{equation}
	\label{eq:rgrav}
	r_{\grav,\I} = 2.4 \times 10^{-7} \left( 1 + \frac{m_\I}{\Msun} \right) \left(\frac{v_\I}{190\,\kms} \right)^{-2} \,\pc \,,
\end{equation}
where `$\I$' represents the impactor.
For reference, the value for the lowest velocities in our simulations is $r_{\grav,\BH} \approx 10^{-3}\,\pc$ and $r_{\grav,\NS} \approx 2 \times 10^{-4}\,\pc$.

\begin{figure*}
	\centering
	\includegraphics[width=\linewidth]{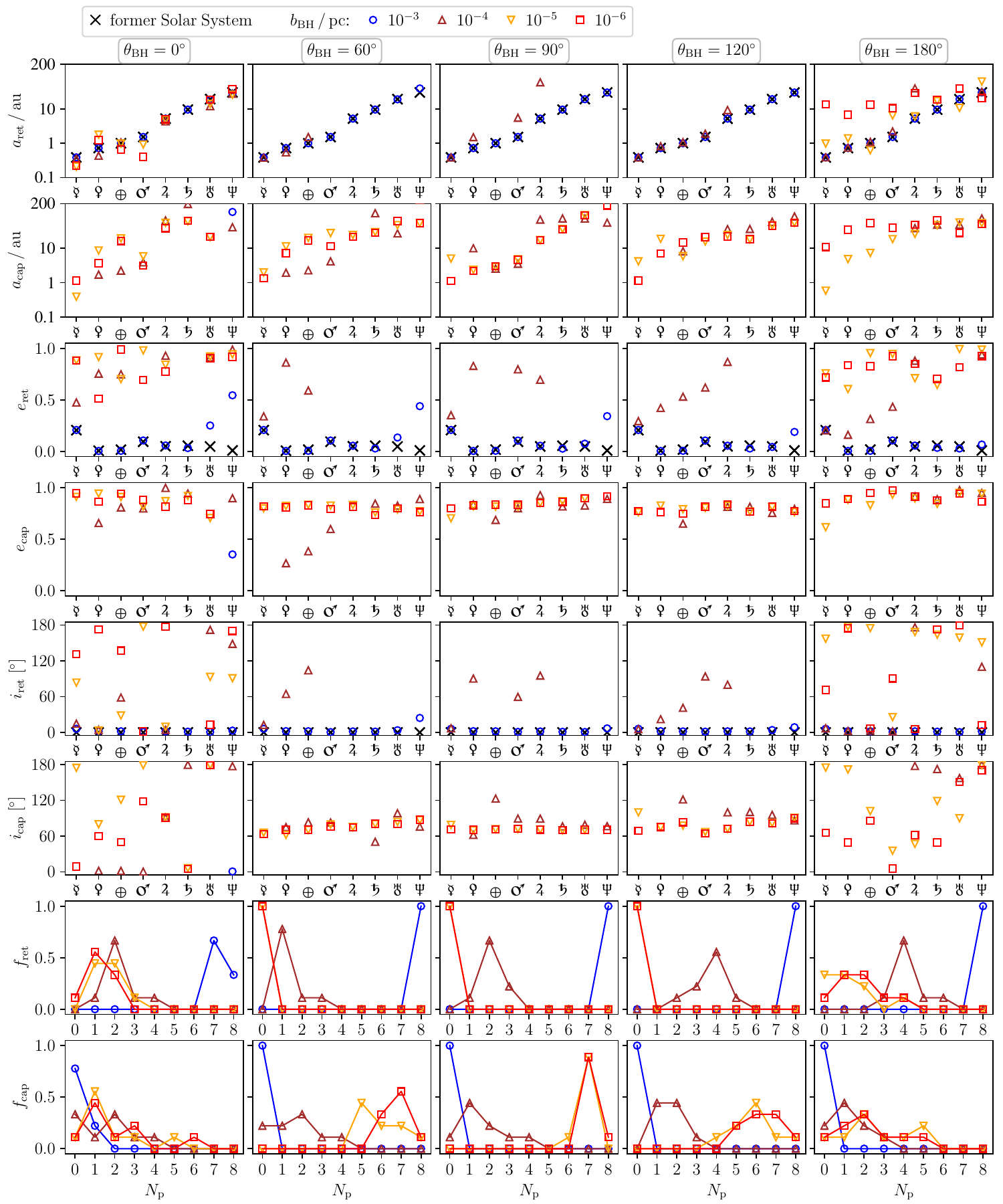}
	\caption{Outcomes of the BH scenario for $v_\BH=10\,\kms$. For each impact parameter (coded by colours and symbols) and incidence angle (columns), we plotted the final distribution of semi-major axes, eccentricities, and inclinations of planets that were either retained by the Sun or captured by the BH (top six rows, for each planet from Mercury~\Mercury\ to Neptune~\Neptune). The bottom two rows show the fraction of retained and captured planets. Each variation in the initial parameters was averaged over nine pre-evolved states of the SS, see Sect.~\ref{sec:methods}. Detailed plots for all $v_\BH$ and $\inc_\BH$ are in Figs.~\ref{fig:bh_cap}--\ref{fig:bh_sun_inc}.}
	\label{fig:bh}
\end{figure*}

\begin{figure*}
	\centering
	\includegraphics[width=\linewidth]{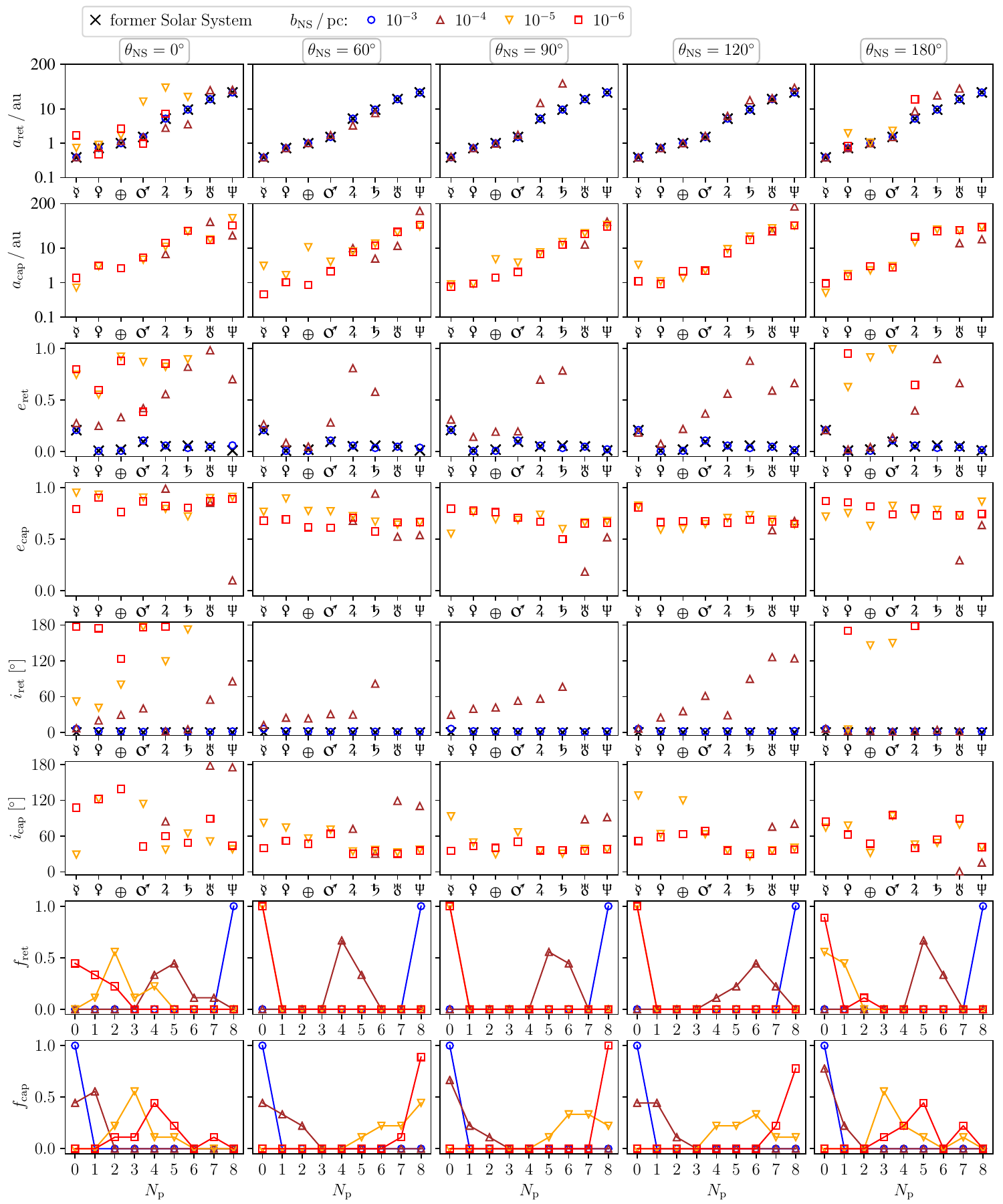}
	\caption{Outcomes of the NS scenario for $v_\NS=10\,\kms$. For each impact parameter (coded by colours and symbols) and incidence angle (columns), we plotted the final distribution of semi-major axes, eccentricities, and inclinations of planets that were either retained by the Sun or captured by the NS (top six rows, for each planet from Mercury~\Mercury\ to Neptune~\Neptune). The bottom two rows show the fraction of retained and captured planets. Each variation in the initial parameters was averaged over nine pre-evolved states of the SS, see Sect.~\ref{sec:methods}. Detailed plots for all $v_\NS$ and $\inc_\NS$ are in Figs.~\ref{fig:ns_cap}--\ref{fig:ns_sun_inc}.}
	\label{fig:ns}
\end{figure*}

For $b_{\BH,\NS} \lesssim r_\grav$, we found alterations in both the inclination and eccentricity of virtually all retained planets regardless of the incidence angle and encounter velocity (see Appendices~\ref{ap:bh} \&~\ref{ap:ns}).
The most efficient captures for both scenarios occurred in the range $60\adeg \lesssim \inc_{\BH,\NS} \lesssim 120\adeg$, with up to eight planets for $b_{\BH,\NS} \lesssim 10^{-5}\,\pc$ and the lowest impact velocity (see $f_\capture$ in Figs.~\ref{fig:bh} \&~\ref{fig:ns}). Eccentricities of the captured planetary orbits, $e_\capture$, were very high -- around 0.8 for the BH and 0.65 for the NS.

There was a~difference between the prograde ($\inc_{\BH,\NS} = 0\adeg$) and retrograde ($\inc_{\BH,\NS} = 180\adeg$) encounters with the SS in both scenarios. A~low-velocity NS showed no directional preferences for planetary capture, but it was most destructive for a~retrograde fly-by with $b_\NS \lesssim 10^{-5}\,\pc$; in 60\,\% of the cases, the SS did not survive and in the remaining 40\,\% the Sun was left with only one terrestrial planet (see $f_\retain$ and $f_\capture$ in Fig.~\ref{fig:ns}).
In contrast, a~low-velocity BH was generally more disruptive and also more efficient in capturing planets for a~prograde encounter (see $f_\retain$ and $f_\capture$ in Fig.~\ref{fig:bh}).

The greatest difference between the NS and BH case was found in the change of the orbital inclinations of the retained planets, $i_\retain$, as the BH is more likely to completely flip the orbits of even massive planets. Based on the positions of the planets during the encounter, both intruders were able to capture planets on prograde, retrograde, or $90\adeg$ orbits while approaching from $\inc_{\BH,\NS} = 0\adeg$ (see $i_\capture$). It was, however, impossible for the NS to capture a~retrograde planet arriving from $\inc_{\NS} = 180\adeg$.

Although the capture process is a~many-body interaction, the final orbital angular momentum and semi-major axis of the captured planet scales directly with the mass of the intruder.
We also evaluated the three-body interaction of the Sun, Earth, and BH. These did not produce the outcomes we found in the complete simulations, as expected \citep[see, e.g.][]{binney_tremaine,heggie_hut}. This three-body interaction is effectively a~two-body encounter because of the large ratios of masses, and the ratio of the impact parameter and the semi-major axis of the Earth's orbit. Consequently, the outcomes of our simulations require the full interaction of the massive planets.

At the smallest impact parameters and lowest velocities, the semi-major axes of the inner retained planets increased more than those of gas giants (see $a_\retain$ in Figs.~\ref{fig:bh} \&~\ref{fig:ns} and Figs.~\ref{fig:bh_semi} \&~\ref{fig:ns_semi}). It should be noted, however, that the most frequent outcome was an ejection of the outer planets.
Unlike previous studies \citep{wang_etal}, we found no hot Jupiters among the retained systems even for the lowest impact parameters for any incident angle or velocity.
We have also compared our NS in-plane encounter with the strong stellar fly-by scenario of \citet{malmberg_etal}. They used a schematic system containing only the giant planets, a $1.5\,\Msun$ intruder, $100\,\au$ for the impact parameter, and $1\,\kms$ as the velocity at infinity. Our closest model is $b_\NS = 10^{-4}\,\pc$ and $v_\NS = 10\,\kms$.
They found that ${\approx}31\,\%$ of their systems ejected a planet immediately after the fly-by (listed in their Table 1) and we found a similar number, $25\,\%$. In contrast, they stated that ${\approx}28\,\%$ of their systems remained stable for up to $100\,\Myr$, while we have no intact systems for any incidence angle. This difference is due to our higher impactor velocity and mass. \citet{malmberg_etal} do not provide the statistics on capture by the intruder, but we see ${\gtrsim}45\,\%$ cases where the NS has captured at least one planet, so the lack of stable systems is partly accounted for by this phenomenon rather than planetary ejection.

\begin{figure*}
	\centering
	NS\hspace{.5\linewidth}
	BH\\
	\includegraphics[width=.495\linewidth]{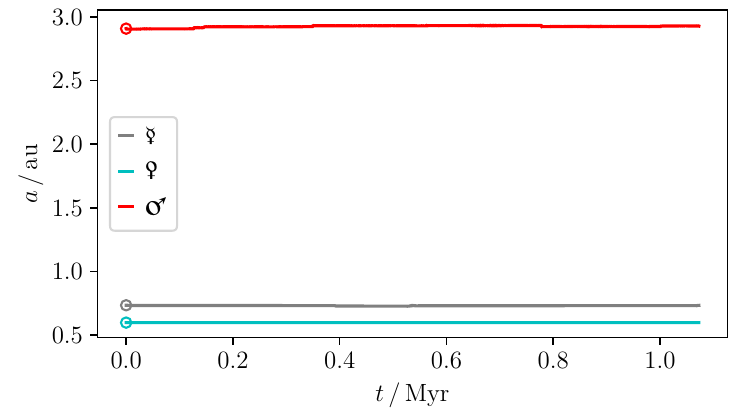}
	\hfill
	\includegraphics[width=.495\linewidth]{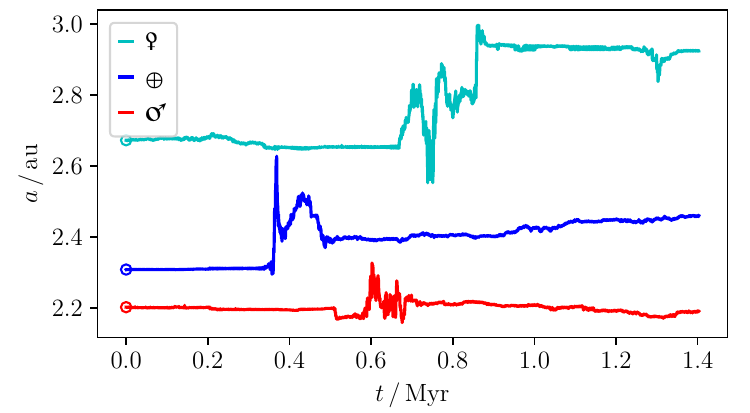}\\
	
	\includegraphics[width=.495\linewidth]{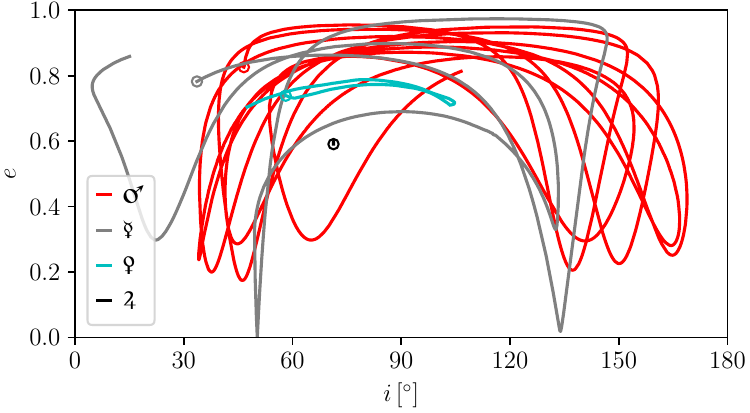}
	\hfill
	\includegraphics[width=.495\linewidth]{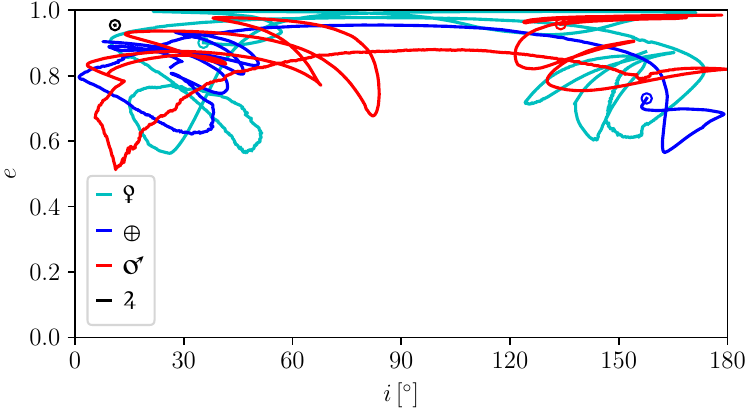}
	\caption{Comparative long-term evolution of captured planetary systems. Left, NS: $b_\NS = 10^{-6}\,\pc$, $\inc_\NS = 90\adeg$, $v_\NS = 50\,\kms$; Right, BH: $b_\BH = 10^{-6}\,\pc$, $\inc_\BH = -16\adeg$, $v_\BH = 10\,\kms$. The capture event is marked by a~circle for each planet. The semi-major axis of Jupiter is not plotted in the top panels, but was $11.3\,\au$ in the NS system and $24.5\,\au$ in the BH system. See text for discussion.} 
	\label{fig:long_term}
\end{figure*}

We also performed two long-term (${>}1\,\Myr$) integrations, one for each impactor, to examine the stability of the captured systems. Both had captured Jupiter and three terrestrial planets. The NS was an orthogonal incidence, while the BH was from the high-resolution study described in Appendix~\ref{ap:bet}. In both cases, all planets were captured in highly eccentric orbits and remained bound to the impactor. The Jovian orbit was stationary while the terrestrial planets displayed the complex behaviour plotted in Fig.~\ref{fig:long_term}. We stress that these orbital histories are randomly chosen examples. Both cases showed long-term inclination oscillations between pro- and retrograde orbits, with the NS system being more regular and it also showed a~larger excursion in eccentricities than the BH case. In the former, Mercury nearly circularised. In contrast, the BH planetary system had a~lower bound in the eccentricity of all planets of ${\approx}0.5$. In the NS case, the semi-major axes remained approximately constant for all planets while in the BH system, the terrestrial planets experienced abrupt jumps that never settled down. This may indicate the onset of longer-term chaotic behaviour. Further discussion is beyond the scope of this letter and a~more extensive analysis will be a~subject of a~future study.

Planetary capture by a~NS is particularly interesting because pulsars have been observed to posses planetary systems (even with multiple planets) -- such as PSR\,1257+12 \citep{1992Natur.355..145W} and PSR\,B1620--26 \citep{1996ASPC..105..525A}. While it is likely that these systems are relics \citep[as, e.g.][]{2006Natur.440..772W}, our models provide an additional formation channel -- the outcome of a~close encounter of a~NS with an ordinary planetary system. The mean rate for such encounters which could lead to a~capture is about $3{\times}10^{-7} (n / \pc^{-3})\,\Myr^{-1}$, where $n$ is the stellar density. For instance, this is ${\gtrsim}0.03\,\Myr^{-1}$ in star clusters which means acquiring a~planet by capture could occur within the active lifetime of a~pulsar (${\lesssim}10^8$ years depending on the period).

\begin{figure*}
	\textbf{a.}\\
	\includegraphics[width=\linewidth]{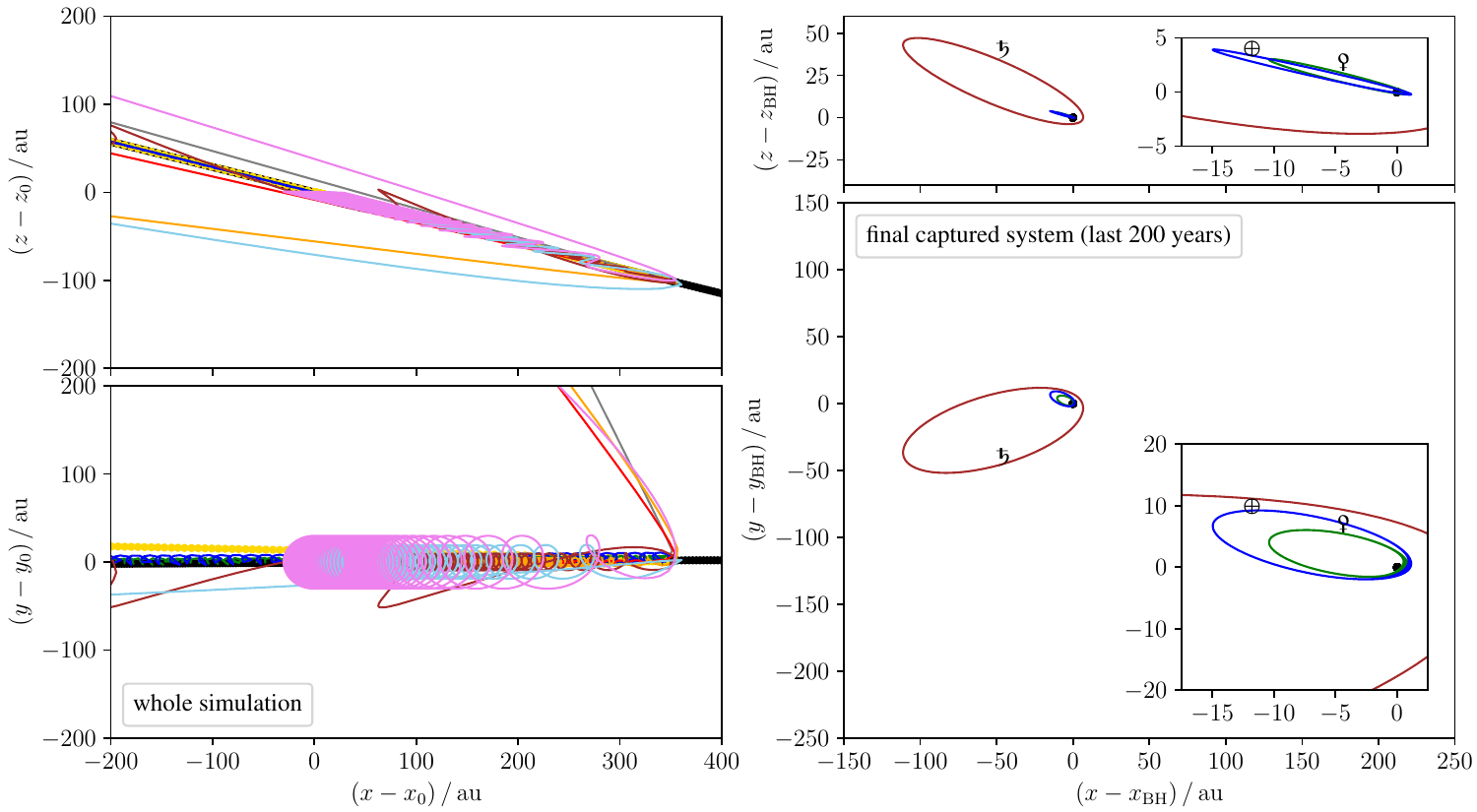}\\
	\textbf{b.}\\
	\includegraphics[width=\linewidth]{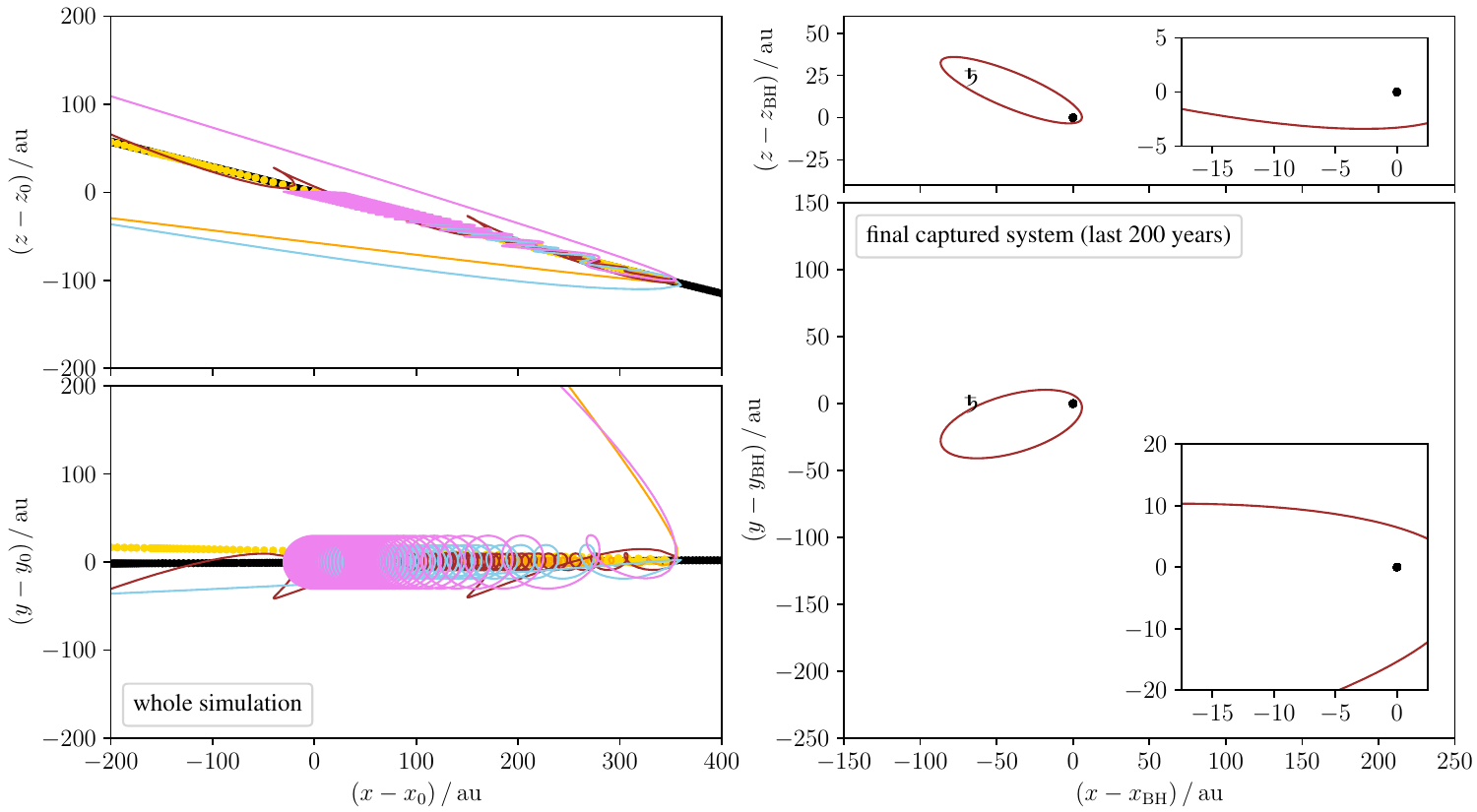}
	\caption{Sample realisation from the high-resolution study (see Appendix~\ref{ap:bet}) with $v_\BH=10\,\kms$ and $b_\BH=10^{-5}\,\pc$; (a) with all planets and (b) without terrestrial planets. Both panels show the trajectories of the planets with coloured lines (\textcolor{Mercury}{Mercury~\Mercury}, \textcolor{Venus}{Venus~\Venus}, \textcolor{Earth}{Earth~\varEarth}, \textcolor{Mars}{Mars~\Mars}, \textcolor{Jupiter}{Jupiter~\Jupiter}, \textcolor{Saturn}{Saturn~\Saturn}, \textcolor{Uranus}{Uranus~\Uranus}, \textcolor{Neptune}{Neptune~\Neptune}), the Sun (yellow dots), and the BH (black dots). The left panels are plotted in the coordinates fixed to the initial position of the Sun ($x_0,y_0,z_0$), in which the impactor arrived from the right-hand side, and the right panels show a~detailed view of the final system captured by the BH in its co-moving coordinates ($x_\BH,y_\BH,z_\BH$).} 
	\label{fig:detail_4-8pl}
\end{figure*}

\begin{acknowledgements}
The project was inspired by a~seminar given at Pisa by E.F.~Guinan, Dec, 2020.
SNS thanks Matteo Cantiello, Mathieu Renzo, and Rubina Kotak for discussions regarding massive stellar evolution and supernovae.
VP greatly appreciates access to computational resources supplied by the project ``e-Infrastruktura CZ'' (e-INFRA LM2018140) provided within the program Projects of Large Research, Development and Innovations Infrastructures.
The NASA ADS, Simbad, and NASA JPL HORIZONS system were used for this research. We thank the anonymous referee for valuable comments.
\end{acknowledgements}

\bibliographystyle{aa}
\bibliography{40454-21}

\begin{thebibliography}{32}
\expandafter\ifx\csname natexlab\endcsname\relax\def\natexlab#1{#1}\fi

\bibitem[{{Arzoumanian} {et~al.}(1996){Arzoumanian}, {Joshi}, {Rasio}, \&
  {Thorsett}}]{1996ASPC..105..525A}
{Arzoumanian}, Z., {Joshi}, K., {Rasio}, F.~A., \& {Thorsett}, S.~E. 1996, in
  Astronomical Society of the Pacific Conference Series, Vol. 105, IAU Colloq.
  160: Pulsars: Problems and Progress, ed. S.~{Johnston}, M.~A. {Walker}, \&
  M.~{Bailes}, 525--530

\bibitem[{{Baumgardt} \& {Sollima}(2017)}]{baumgardt_sollima}
{Baumgardt}, H. \& {Sollima}, S. 2017, \mnras, 472, 744

\bibitem[{{Belczynski} {et~al.}(2008){Belczynski}, {Kalogera}, {Rasio}, {Taam},
  {Zezas}, {Bulik}, {Maccarone}, \& {Ivanova}}]{belczynski}
{Belczynski}, K., {Kalogera}, V., {Rasio}, F.~A., {et~al.} 2008, \apjs, 174,
  223

\bibitem[{{Berski} \& {Dybczy{\'n}ski}(2016)}]{gliese_710}
{Berski}, F. \& {Dybczy{\'n}ski}, P.~A. 2016, \aap, 595, L10

\bibitem[{Binney \& Tremaine(2008)}]{binney_tremaine}
Binney, J. \& Tremaine, S. 2008, Galactic Dynamics: Second Edition, Princeton
  Series in Astrophysics (Princeton University Press)

\bibitem[{{Cai} {et~al.}(2017){Cai}, {Kouwenhoven}, {Portegies Zwart}, \&
  {Spurzem}}]{2017MNRAS.470.4337C}
{Cai}, M.~X., {Kouwenhoven}, M.~B.~N., {Portegies Zwart}, S.~F., \& {Spurzem},
  R. 2017, \mnras, 470, 4337

\bibitem[{{Dehnen} \& {Binney}(1998)}]{lsr_hipparcos}
{Dehnen}, W. \& {Binney}, J.~J. 1998, \mnras, 298, 387

\bibitem[{{Dybczy{\'n}ski}(2006)}]{oort_cloud}
{Dybczy{\'n}ski}, P.~A. 2006, \aap, 449, 1233

\bibitem[{{Famaey} {et~al.}(2005){Famaey}, {Jorissen}, {Luri}, {Mayor}, {Udry},
  {Dejonghe}, \& {Turon}}]{betelgeuse_vel}
{Famaey}, B., {Jorissen}, A., {Luri}, X., {et~al.} 2005, \aap, 430, 165

\bibitem[{{Fragione} {et~al.}(2018){Fragione}, {Pavl{\'\i}k}, \&
  {Banerjee}}]{msp}
{Fragione}, G., {Pavl{\'\i}k}, V., \& {Banerjee}, S. 2018, \mnras, 480, 4955

\bibitem[{{Fryer} {et~al.}(2012){Fryer}, {Belczynski}, {Wiktorowicz},
  {Dominik}, {Kalogera}, \& {Holz}}]{fryer}
{Fryer}, C.~L., {Belczynski}, K., {Wiktorowicz}, G., {et~al.} 2012, \apj, 749

\bibitem[{{Hansen} \& {Phinney}(1997)}]{kicks_ns}
{Hansen}, B.~M.~S. \& {Phinney}, E.~S. 1997, \mnras, 291, 569

\bibitem[{{Hao} {et~al.}(2013){Hao}, {Kouwenhoven}, \&
  {Spurzem}}]{2013MNRAS.433..867H}
{Hao}, W., {Kouwenhoven}, M.~B.~N., \& {Spurzem}, R. 2013, \mnras, 433, 867

\bibitem[{Harris {et~al.}(2020)Harris, Millman, van~der Walt, Gommers,
  Virtanen, Cournapeau, Wieser, Taylor, Berg, Smith, Kern, Picus, Hoyer, van
  Kerkwijk, Brett, Haldane, del R{'{\i}}o, Wiebe, Peterson,
  G{'{e}}rard-Marchant, Sheppard, Reddy, Weckesser, Abbasi, Gohlke, \&
  Oliphant}]{numpy}
Harris, C.~R., Millman, K.~J., van~der Walt, S.~J., {et~al.} 2020, Nature, 585,
  357

\bibitem[{Heggie \& Hut(2003)}]{heggie_hut}
Heggie, D.~C. \& Hut, P. 2003, The Gravitational Million-Body Problem:
  A~Multidisciplinary Approach to Star Cluster Dynamics (Cambridge, UK:
  Cambridge University Press)

\bibitem[{Hunter(2007)}]{matplotlib}
Hunter, J.~D. 2007, Computing in Science \& Engineering, 9, 90

\bibitem[{{Jonker} \& {Nelemans}(2004)}]{bh_ns_kicks}
{Jonker}, P.~G. \& {Nelemans}, G. 2004, \mnras, 354, 355

\bibitem[{{Laughlin} \& {Adams}(2000)}]{laughlin_adams}
{Laughlin}, G. \& {Adams}, F.~C. 2000, \icarus, 145, 614

\bibitem[{{Li} \& {Adams}(2015)}]{li_adams}
{Li}, G. \& {Adams}, F.~C. 2015, \mnras, 448, 344

\bibitem[{{Malmberg} {et~al.}(2011){Malmberg}, {Davies}, \&
  {Heggie}}]{malmberg_etal}
{Malmberg}, D., {Davies}, M.~B., \& {Heggie}, D.~C. 2011, \mnras, 411, 859

\bibitem[{{Pavl{\'\i}k} {et~al.}(2018){Pavl{\'\i}k}, {Je{\v{r}}{\'a}bkov{\'a}},
  {Kroupa}, \& {Baumgardt}}]{pavl_bh}
{Pavl{\'\i}k}, V., {Je{\v{r}}{\'a}bkov{\'a}}, T., {Kroupa}, P., \& {Baumgardt},
  H. 2018, \aap, 617, A69

\bibitem[{{Peuten} {et~al.}(2016){Peuten}, {Zocchi}, {Gieles}, {Gualandris}, \&
  {H{\'e}nault-Brunet}}]{peuten}
{Peuten}, M., {Zocchi}, A., {Gieles}, M., {Gualandris}, A., \&
  {H{\'e}nault-Brunet}, V. 2016, \mnras, 462, 2333

\bibitem[{{Rein} \& {Liu}(2012)}]{rebound}
{Rein}, H. \& {Liu}, S.~F. 2012, \aap, 537, A128

\bibitem[{{Rein} \& {Spiegel}(2015)}]{reboundias15}
{Rein}, H. \& {Spiegel}, D.~S. 2015, \mnras, 446, 1424

\bibitem[{{Repetto} {et~al.}(2017){Repetto}, {Igoshev}, \&
  {Nelemans}}]{Repetto2017}
{Repetto}, S., {Igoshev}, A.~P., \& {Nelemans}, G. 2017, \mnras, 467, 298

\bibitem[{{Shara} {et~al.}(2016){Shara}, {Hurley}, \&
  {Mardling}}]{2016ApJ...816...59S}
{Shara}, M.~M., {Hurley}, J.~R., \& {Mardling}, R.~A. 2016, \apj, 816, 59

\bibitem[{{Spurzem} {et~al.}(2009){Spurzem}, {Giersz}, {Heggie}, \&
  {Lin}}]{2009ApJ...697..458S}
{Spurzem}, R., {Giersz}, M., {Heggie}, D.~C., \& {Lin}, D.~N.~C. 2009, \apj,
  697, 458

\bibitem[{Varri {et~al.}(2018)Varri, Cai, Concha-Ram{\'i}rez, Dinnbier,
  L{\"u}tzgendorf, Pavl{\'i}k, Rastello, Sollima, Wang, \& Zocchi}]{Varri2018}
Varri, A.~L., Cai, M.~X., Concha-Ram{\'i}rez, F., {et~al.} 2018, Computational
  Astrophysics and Cosmology, 5, 2

\bibitem[{{Wang} {et~al.}(2020){Wang}, {Leigh}, {Perna}, \&
  {Shara}}]{wang_etal}
{Wang}, Y.-H., {Leigh}, N. W.~C., {Perna}, R., \& {Shara}, M.~M. 2020, \apj,
  905, 136

\bibitem[{{Wang} {et~al.}(2006){Wang}, {Chakrabarty}, \&
  {Kaplan}}]{2006Natur.440..772W}
{Wang}, Z., {Chakrabarty}, D., \& {Kaplan}, D.~L. 2006, \nat, 440, 772

\bibitem[{{Wolszczan} \& {Frail}(1992)}]{1992Natur.355..145W}
{Wolszczan}, A. \& {Frail}, D.~A. 1992, \nat, 355, 145

\bibitem[{{Zheng} {et~al.}(2015){Zheng}, {Kouwenhoven}, \&
  {Wang}}]{2015MNRAS.453.2759Z}
{Zheng}, X., {Kouwenhoven}, M.~B.~N., \& {Wang}, L. 2015, \mnras, 453, 2759

\end{thebibliography}

\clearpage
\appendix

\section{High-resolution BH simulations}
\label{ap:bet}

Our study was initiated asking what would happen if a~BH, expected to form in the Betelgeuse SN, received a~kick that would direct it towards the Solar System. Exploring this required specific initial conditions linked to the progenitor and a~high time resolution study. We found that the lower resolution generic cases that bounded these conditions were consistent.

For what we hereafter call `the Betelgeuse case', the remnant had $m_\alpha = 10\,\Msun$ and was launched from the ecliptic latitude of the star, $-16.027\adeg$ (J2000.0). We used this as the incidence angle in all of the high-resolution simulations. In the following discussion, we use the subscript $\alpha$ for `$\I$' to label all the parameters.
We used encounter velocities $10\,\kms \leq v_\alpha \leq 190\,\kms$. These were set up in the barycentric coordinates of the SS and, therefore, are not the actual kick velocity of \aOri\ which has to be corrected for the stellar radial velocity \citep{betelgeuse_vel}, that is, the kick would be $\approx 22\,\kms$ higher.

\begin{figure*}
	\centering
	\includegraphics[width=\linewidth]{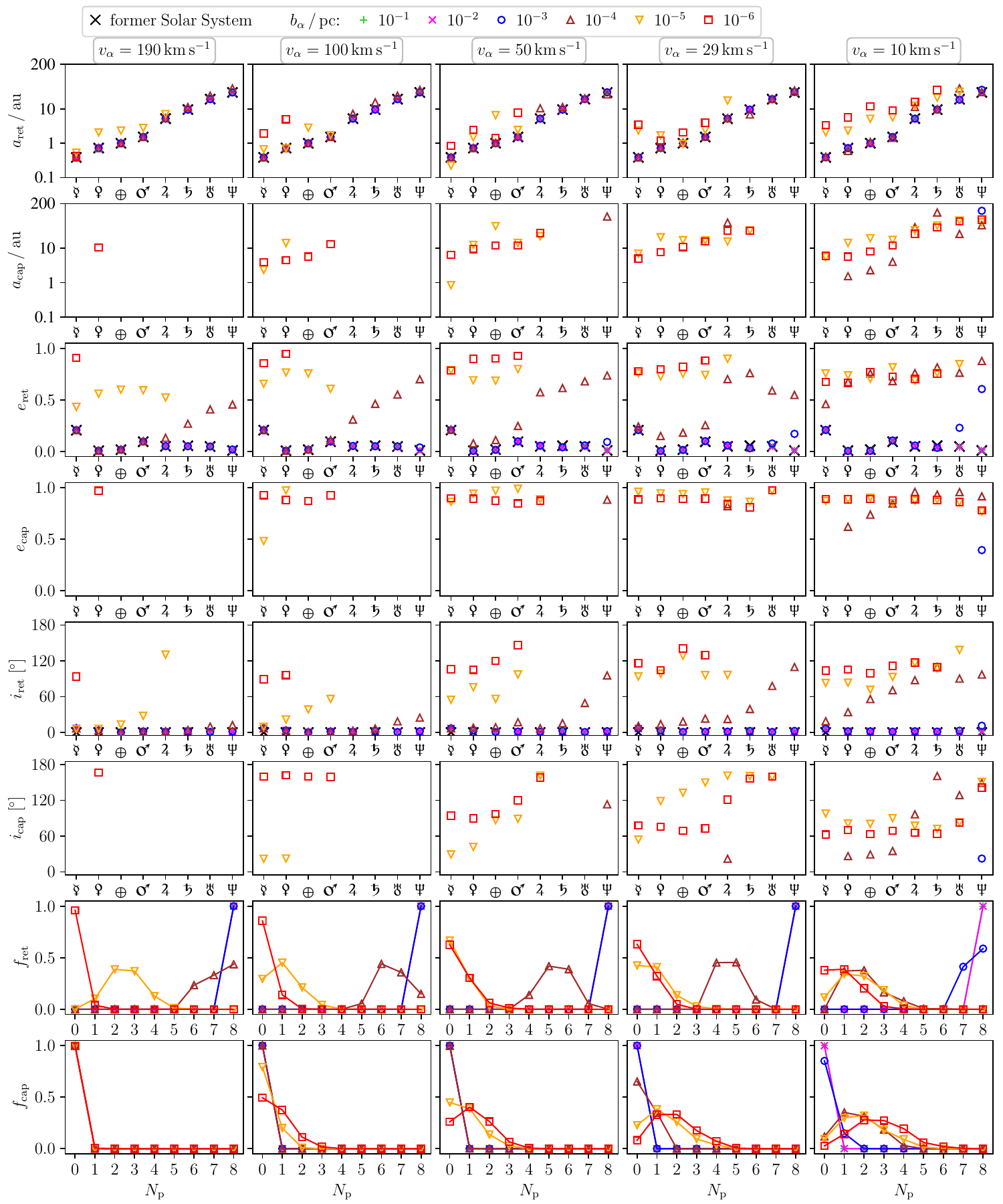}
	\caption{Outcomes of the Betelgeuse scenario. For each impact parameter (coded by colours and symbols) and encounter velocity (columns), we plotted the final distribution of semi-major axes, eccentricities, and inclinations of planets that were either retained by the Sun or captured by \aOri\ (top six rows, for each planet from Mercury~\Mercury\ to Neptune~\Neptune). The bottom two rows show the fraction of retained and captured planets. All data were averaged over 330 simulations.}
	\label{fig:bet}
\end{figure*}

Our principal results are shown in Fig.~\ref{fig:bet} and a~sample realisation is shown in Fig.~\ref{fig:detail_4-8pl}a. For $b_\alpha > 10^{-2}\,\pc$, the SS remained stable regardless of $v_\alpha$ (see $f_\retain$). The planetary orbits remained unchanged except for Mercury, whose inclination increased to $5\adeg$ (see $a_\retain$, $e_\retain$, $i_\retain$).
At these distances, the \aOri\ remnant did not significantly alter the trajectory of the SS in the Galaxy -- the maximum deviation of $0.8\,\kms$ (i.e. ${\approx}10\,\%$ of the peculiar velocity of the Sun) was obtained for $v_\alpha = 10\,\kms$ -- with a~decreasing impact parameter, this effect increased (see Tab.~\ref{tab:vel}).

\setlength\tabcolsep{3.5pt}
\begin{table}[!b]
	\centering
	\caption{Deviation (in absolute values) from the peculiar velocity of the Sun \citep{lsr_hipparcos}, i.e. $U \approx 10.00\,\kms$ (towards the Galactic centre), $V \approx 5.25\,\kms$ (in the orbital motion of the Sun), and $W \approx 7.17\,\kms$ (perpendicular to the Galactic plane).}
	\begin{tabular}{ccrrrrr}
		\hline
		\multicolumn{2}{c}{$v_\alpha\,/\,\kms\rightarrow$} &	190	&	100	&	50	&	29	&	10 \\
		$\downarrow\,b_\alpha\,/\,\pc$		& $\downarrow\kms$ &&&&&\\
		\hline
		$10^{-1}$	 & $\Delta U$ &   0.00 &   0.00 &  0.01 &  0.01 &  0.01 \\
							 & $\Delta V$ &   0.00 &   0.00 &  0.01 &  0.02 &  0.05 \\
							 & $\Delta W$ &   0.00 &   0.00 &  0.00 &  0.00 &  0.00 \\[5pt] 
		                                                    
		$10^{-2}$	 & $\Delta U$ &   0.02 &   0.03 &  0.06 &  0.10 &  0.32 \\
							 & $\Delta V$ &   0.04 &   0.08 &  0.16 &  0.27 &  0.80 \\ 
							 & $\Delta W$ &   0.00 &   0.00 &  0.00 &  0.00 &  0.03 \\[5pt]
		                                                    
		$10^{-3}$	 & $\Delta U$ &   0.15 &   0.29 &  0.61 &  1.08 &  4.77 \\
							 & $\Delta V$ &   0.42 &   0.81 &  1.61 &  2.66 &  5.78 \\
							 & $\Delta W$ &   0.00 &   0.00 &  0.02 &  0.10 &  2.20 \\[5pt]
		                                     
		$10^{-4}$	 & $\Delta U$ &   1.58 &   3.20 &  7.86 & 16.02 & 13.50 \\
							 & $\Delta V$ &   4.25 &   7.98 & 14.84 & 18.13 &  1.01 \\
							 & $\Delta W$ &   0.04 &   0.27 &  2.07 &  7.77 & 11.40 \\[5pt]
		                           
		$10^{-5}$	 & $\Delta U$ &  19.23 &  47.47 & 63.19 & 40.76 & 13.00 \\
							 & $\Delta V$ &  40.43 &  57.68 & 17.15 &  4.04 &  4.28 \\
							 & $\Delta W$ &   3.83 &  21.87 & 46.66 & 34.55 & 11.90 \\[5pt]
		
		$10^{-6}$	 & $\Delta U$ & 211.61 & 136.01 & 65.98 & 39.06 & 12.90 \\
							 & $\Delta V$ & 100.80 &   9.95 & 18.65 & 12.95 &  4.61 \\
							 & $\Delta W$ & 143.42 & 114.25 & 59.49 & 35.78 & 11.90 \\
		\hline
	\end{tabular}
	\label{tab:vel}
\end{table}
\setlength\tabcolsep{6pt}

For $10^{-2} \gtrsim b_\alpha / \pc \gtrsim 10^{-3}$, the Sun still retained all planets if the remnant's velocity was at least as great as the Earth's orbital velocity. If it is very slow, $v_\alpha=10\,\kms$, it takes ${\sim}100\,\yr$ to traverse the SS, which is enough time to significantly alter the eccentricities of the two outermost planets. This either unbinds Neptune or results in its capture.

For the highest velocity encounters, no planets were captured at any impact parameter above $10^{-6}\,\pc$. For all impact parameters $b_\alpha < 10^{-3}\,\pc$, the SS was considerably altered, especially regarding the number of planets retained and their orbital eccentricities, as shown in Fig.~\ref{fig:bet} ($f_\retain$ and $e_\retain$). There were no discernible effects above impact parameters $10^{-3}\,\pc$, except for the $10\,\kms$ encounter velocity.

\begin{figure}
	\centering
	\includegraphics[width=.9\linewidth]{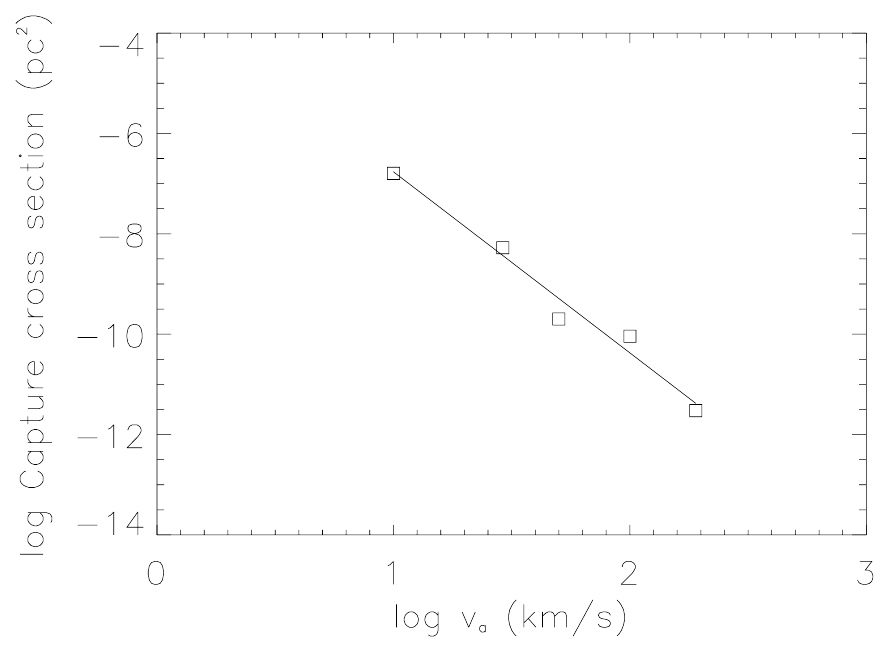}
	\caption{Frequency weighted capture cross section for the Betelgeuse scenario. The data points were calculated from our simulations. See text for details.}
	\label{fig:cross-fit}
\end{figure}

The capture cross section resulting from the simulations including the entire system is $\Sigma_\capture \lesssim 3{\times}10^{-6}\,\pc^2$, which agrees with the simple estimate from Eq.~\eqref{eq:rgrav}. The number of captured planets depended sensitively on $v_\alpha$.
At $10\,\kms$, all of the encounters with $b_\alpha \lesssim 10^{-3}\,\pc$ resulted in a~capture of at least one planet in more than $15\,\%$ of the simulations.
In contrast, at $50\,\kms$, only 1/330 simulations resulted in the capture of a~single planet (Neptune, see Fig.~\ref{fig:bet}) with the maximum cross section $\Sigma_\capture \approx 3{\times}10^{-8}\,\pc^2$.  The number of planets captured was $\Np \leq 7$, all of which had high orbital eccentricities and inclinations. Regarding disruptions, at least 55\,\% of the simulations with $b_\alpha \leq 10^{-4}$ resulted in the ejection of at least one planet regardless of $v_\alpha$.
We stress that these values are frequencies, not probabilities, because our models are not Monte Carlo simulations. As shown in Fig.~\ref{fig:cross-fit}, we find a~scaling relation for the \aOri\ capture cross section weighted by the frequency of models producing at least one capture of
\begin{equation}
        \Sigma_{\capture,\alpha} \approx 7{\times}10^{-4} \left( v_\alpha \,\big/\, \kms \right)^{-3.6}\,\pc^2 \,.
\end{equation}

As a~final point regarding the Betelgeuse scenario, although a~disruptive encounter of its remnant with the SS is highly unlikely, even a~more probable distant encounter would not be particularly pleasant.
The models we present did not include any minor bodies (not even planetary moons). Their masses are negligible relative to the planets or the Sun, so they have no effect on the dynamical stability of the SS, but it is certain that the close passage of a~massive body will alter their positions and velocities. 
We performed a~few simulations with Halley-type comets ($a \sim 200\,\au$), but even for an impact parameter as low as $10^{-2}\,\pc$, their orbits in the SS barycentric frame were only marginally altered.
Nonetheless, a~distant passage of a~massive object, $0.1{-}1\,\pc$, could strip away part of the Oort cloud and also produce an excess of comets with very long periods and high eccentricities that could potentially cross Earth's orbit, as found in previous studies of stellar fly-bys \citep{oort_cloud,gliese_710}.

\onecolumn
\section{Additional figures: BH simulations}
\label{ap:bh}

\begin{figure*}[!htb]
	\centering
	\includegraphics[width=.6\linewidth]{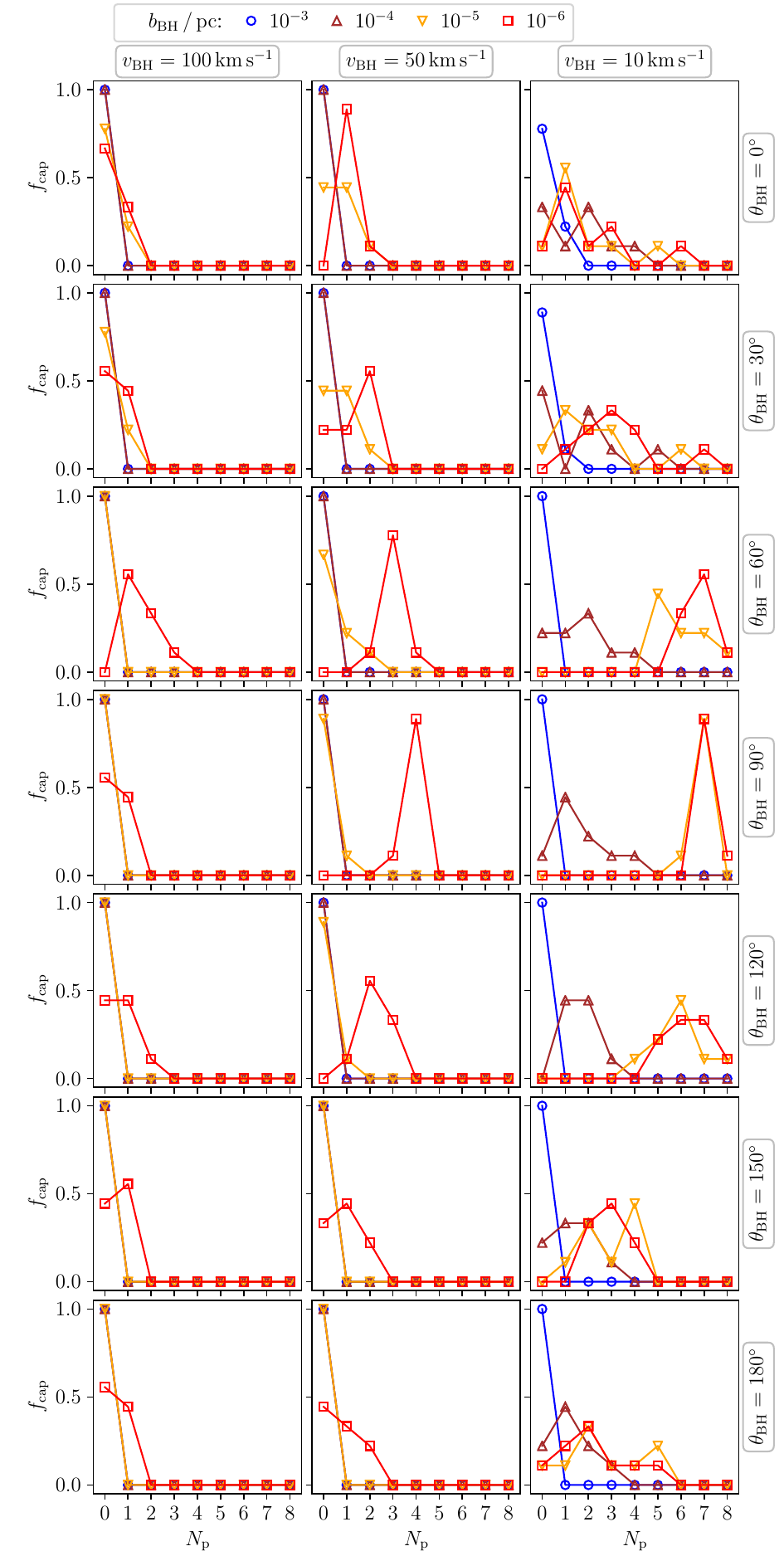}
	\caption{Capture of planets by a~BH. Frequency of capture, $f_\capture$, of a~given number of planets, $\Np$, by the BH for the specified set of initial conditions (impact parameter, $b_\BH$, encounter velocity, $v_\BH$, and the incidence angle, $\inc_\BH$).}
	\label{fig:bh_cap}
\end{figure*}

\begin{figure*}
	\centering
	\includegraphics[width=.6\linewidth]{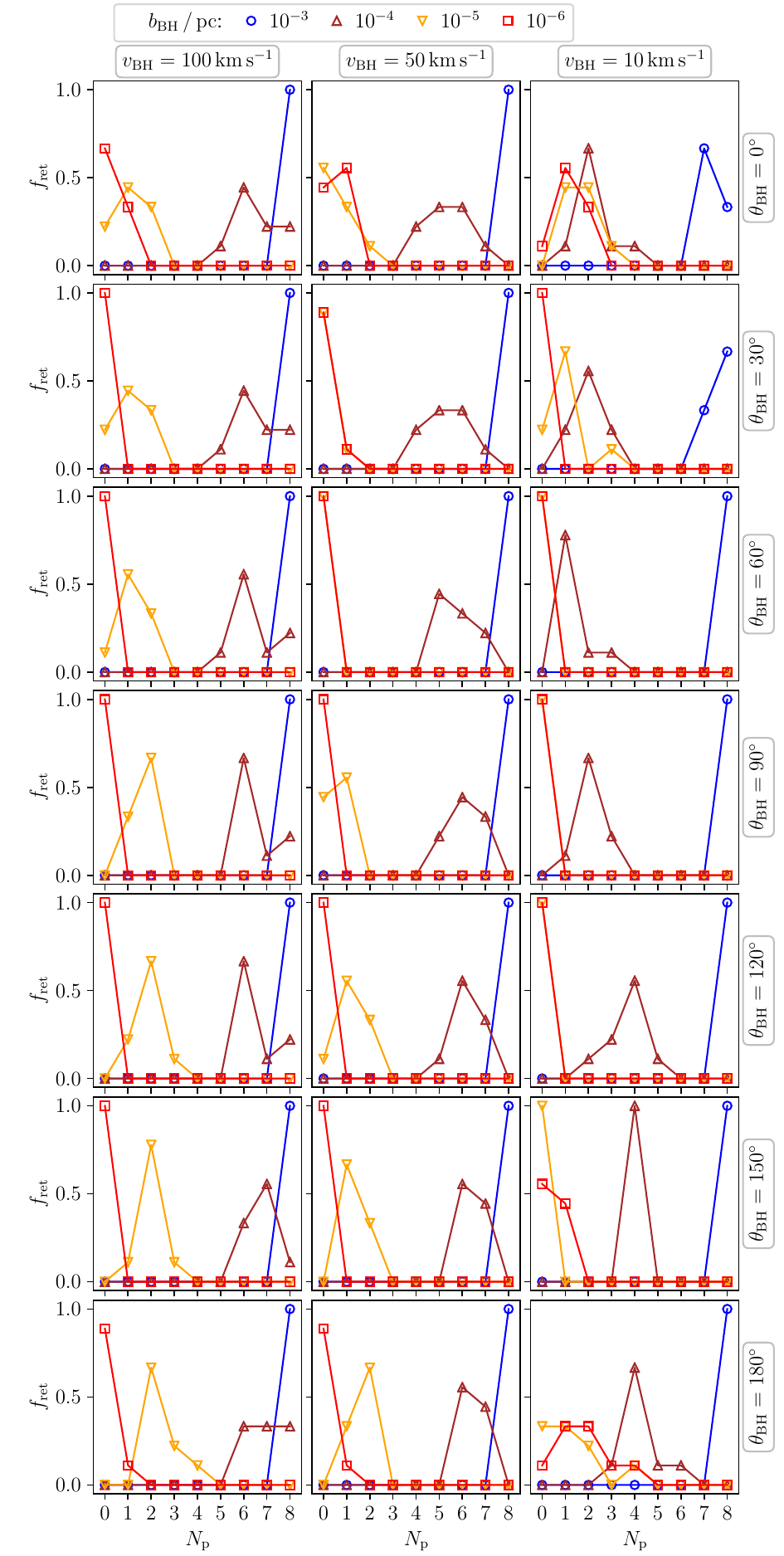}
	\caption{Planet retention after a~BH encounter. Frequency of retention, $f_\retain$, of a~given number of planets, $\Np$, by the Sun after the BH fly-by for the specified set of initial conditions (impact parameter, $b_\BH$, encounter velocity, $v_\BH$, and the incidence angle, $\inc_\BH$).}
	\label{fig:bh_sun_ret}
\end{figure*}

\begin{figure*}
	\centering
	\includegraphics[width=.6\linewidth]{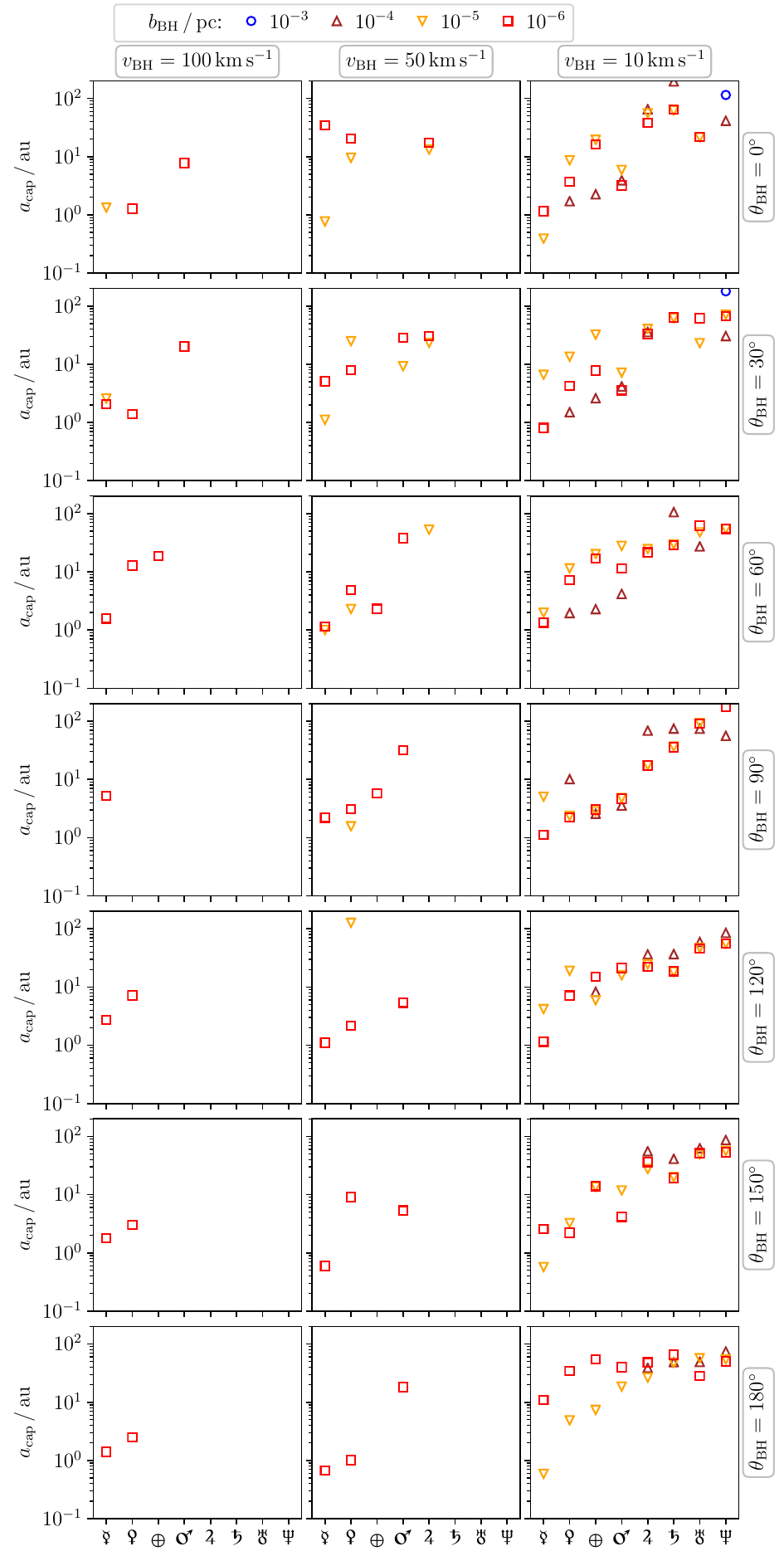}
	\caption{Distribution of semi-major axes of planets captured by a~BH. Each data point represents the mean value of the semi-major axis of a~specific planet (Mercury~\Mercury\ to Neptune~\Neptune) if it was captured by a~BH. The data were compiled from all pre-evolved states of the SS for a~given set of initial conditions (impact parameter, $b_\BH$, encounter velocity, $v_\BH$, and the incidence angle, $\inc_\BH$) -- i.e. the panels do not show how many planets were captured in total (see Fig.~\ref{fig:bh_cap} for reference).}
	\label{fig:bh_semi}
\end{figure*}

\begin{figure*}
	\centering
	\includegraphics[width=.6\linewidth]{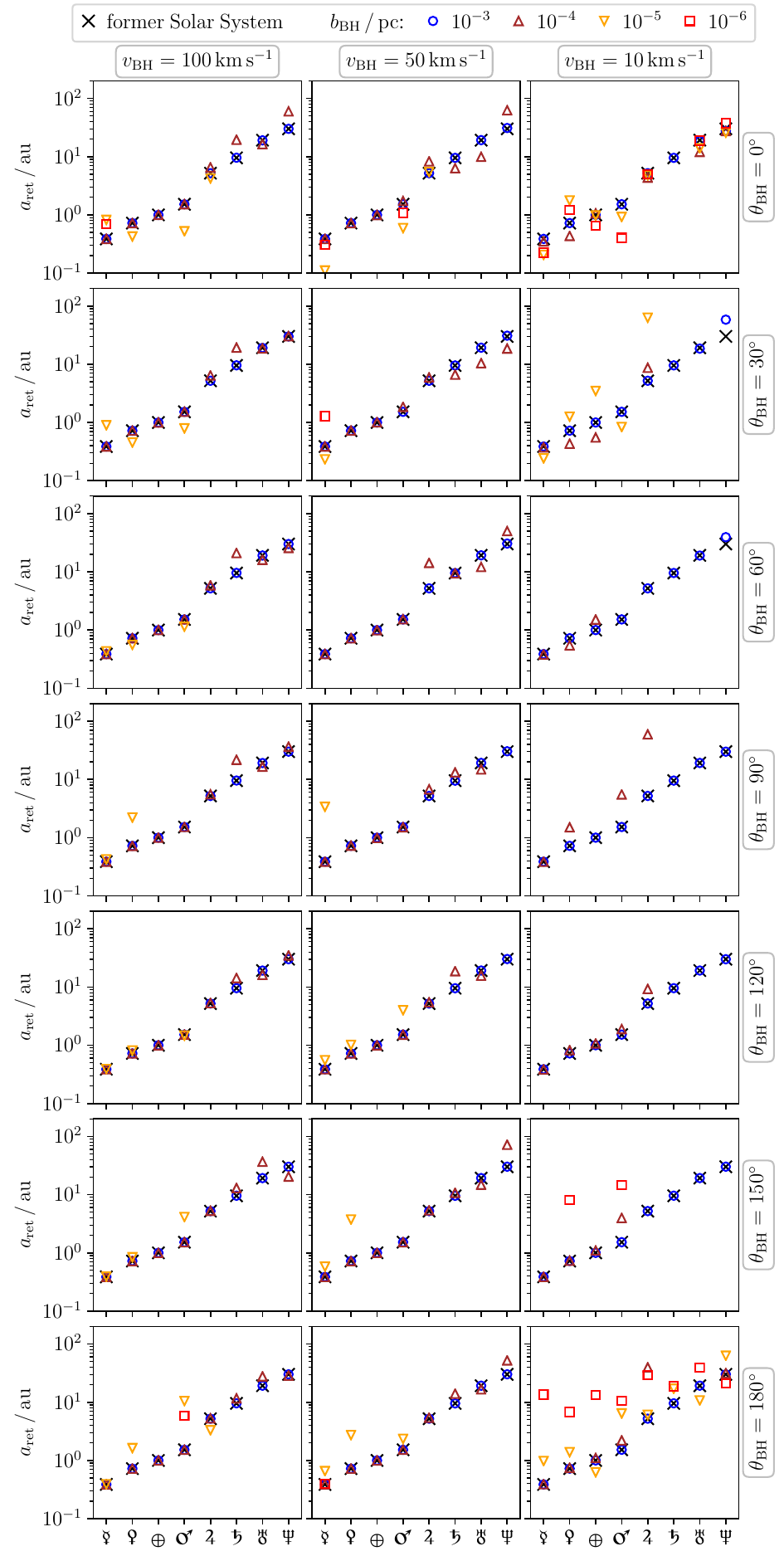}
	\caption{Distribution of semi-major axes of planets retained by the Sun after a~BH encounter. Each data point represents the mean value of the semi-major axis of a~specific planet (Mercury~\Mercury\ to Neptune~\Neptune) if it was retained by the Sun. The data were compiled from all pre-evolved states of the SS for a~given set of initial conditions (impact parameter, $b_\BH$, encounter velocity, $v_\BH$, and the incidence angle, $\inc_\BH$) -- i.e. the panels do not show how many planets were retained in total (see Fig.~\ref{fig:bh_sun_ret}). The former SS is shown for reference.}
	\label{fig:bh_sun_semi}
\end{figure*}

\begin{figure*}
	\centering
	\includegraphics[width=.6\linewidth]{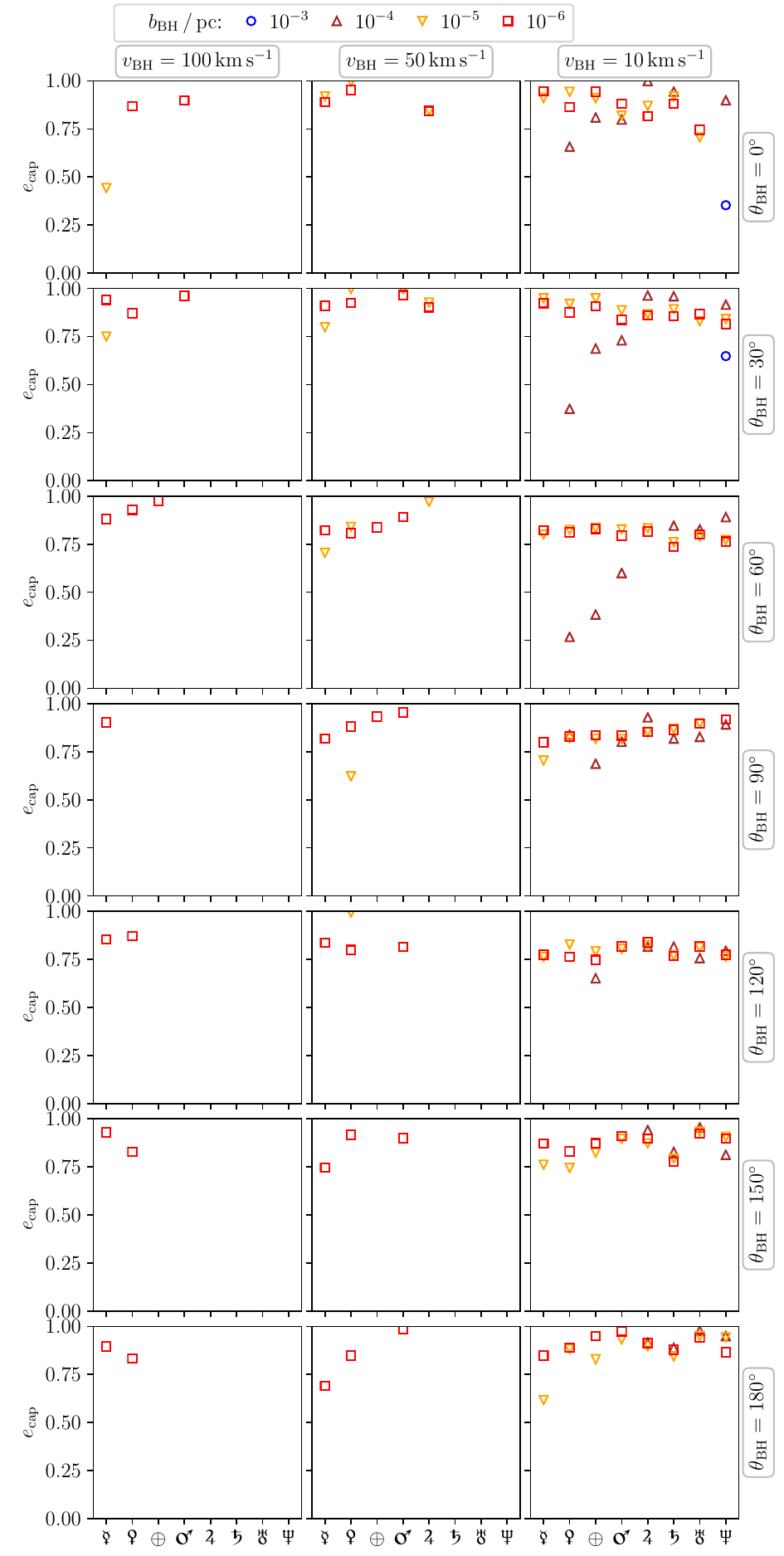}
	\caption{Distribution of eccentricities of planets captured by a~BH. Each data point represents the mean value of the eccentricity of a~specific planet (Mercury~\Mercury\ to Neptune~\Neptune) if it was captured by a~BH. The data were compiled from all pre-evolved states of the SS for a~given set of initial conditions (impact parameter, $b_\BH$, encounter velocity, $v_\BH$, and the incidence angle, $\inc_\BH$) -- i.e. the panels do not show how many planets were captured in total (see Fig.~\ref{fig:bh_cap} for reference).}
	\label{fig:bh_ecc}
\end{figure*}

\begin{figure*}
	\centering
	\includegraphics[width=.6\linewidth]{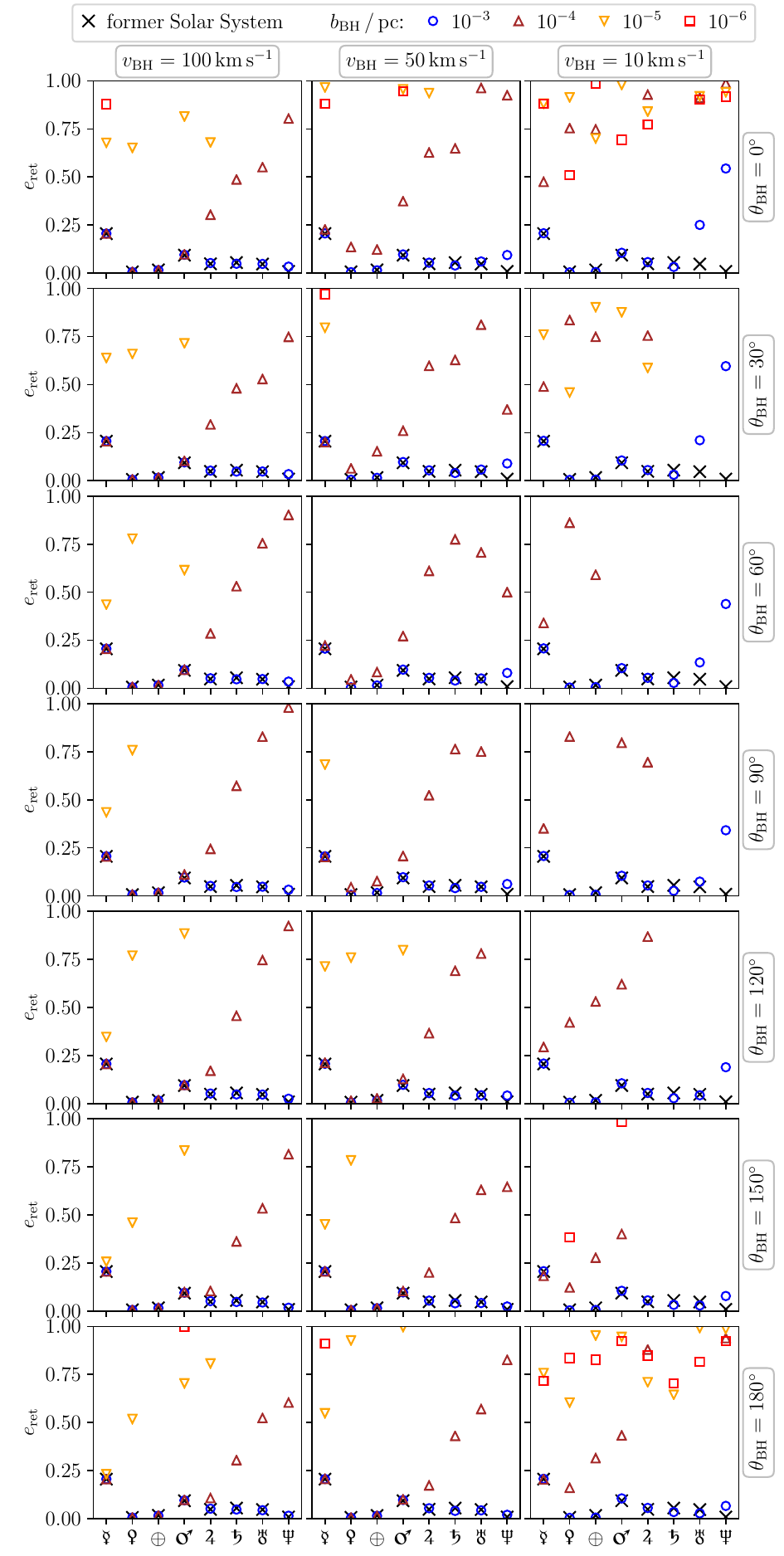}
	\caption{Distribution of eccentricities of planets retained by the Sun after a~BH encounter. Each data point represents the mean value of the eccentricity of a~specific planet (Mercury~\Mercury\ to Neptune~\Neptune) if it was retained by the Sun. The data were compiled from all pre-evolved states of the SS for a~given set of initial conditions (impact parameter, $b_\BH$, encounter velocity, $v_\BH$, and the incidence angle, $\inc_\BH$) -- i.e. the panels do not show how many planets were retained in total (see Fig.~\ref{fig:bh_sun_ret}). The former SS is shown for reference.}
	\label{fig:bh_sun_ecc}
\end{figure*}

\begin{figure*}
	\centering
	\includegraphics[width=.6\linewidth]{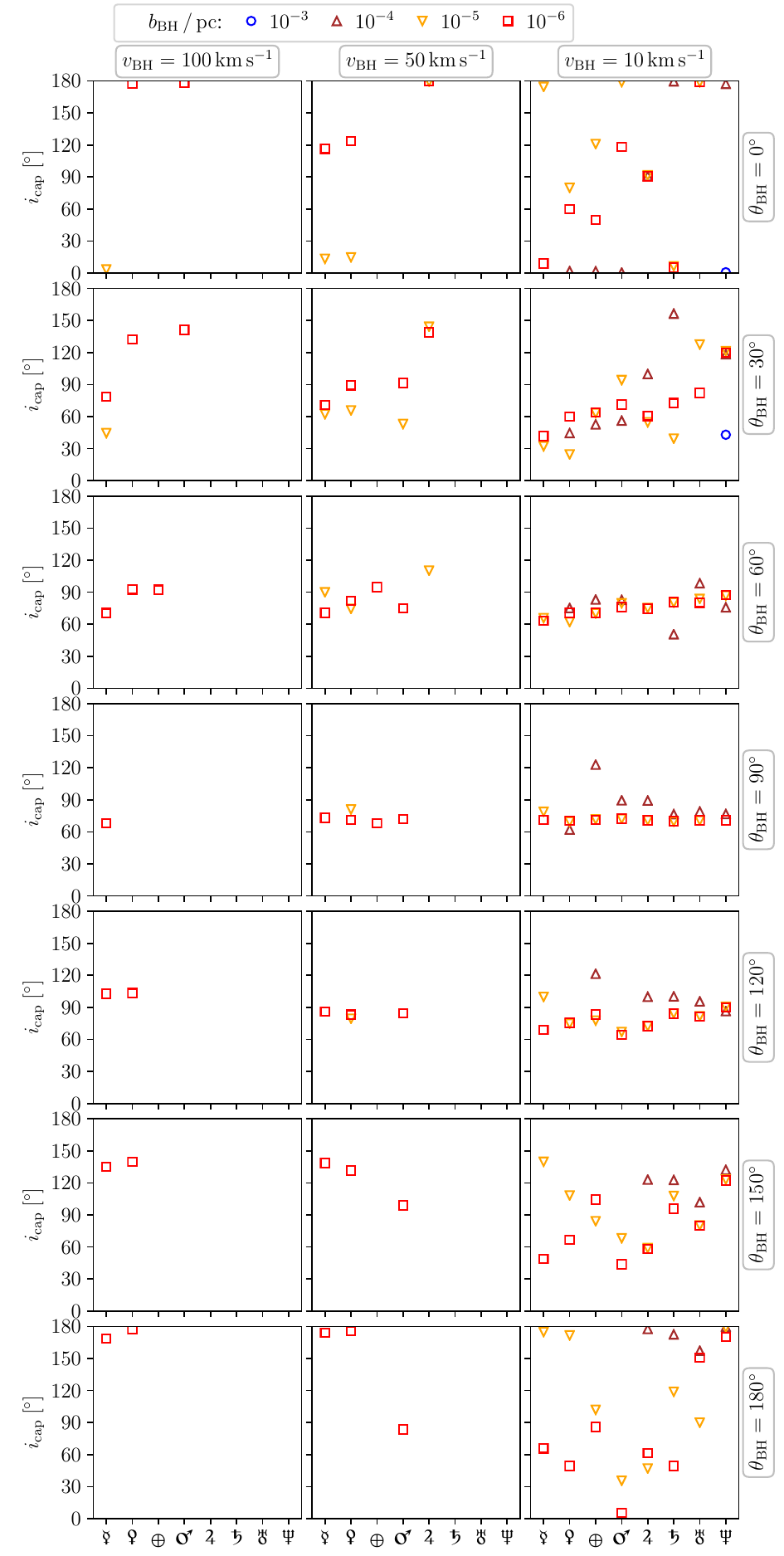}
	\caption{Distribution of inclinations of planets captured by a~BH. Each data point represents the mean value of the inclination of a~specific planet (Mercury~\Mercury\ to Neptune~\Neptune) if it was captured by a~BH. The data were compiled from all pre-evolved states of the SS for a~given set of initial conditions (impact parameter, $b_\BH$, encounter velocity, $v_\BH$, and the incidence angle, $\inc_\BH$) -- i.e. the panels do not show how many planets were captured in total (see Fig.~\ref{fig:bh_cap} for reference).}
	\label{fig:bh_inc}
\end{figure*}

\begin{figure*}
	\centering
	\includegraphics[width=.6\linewidth]{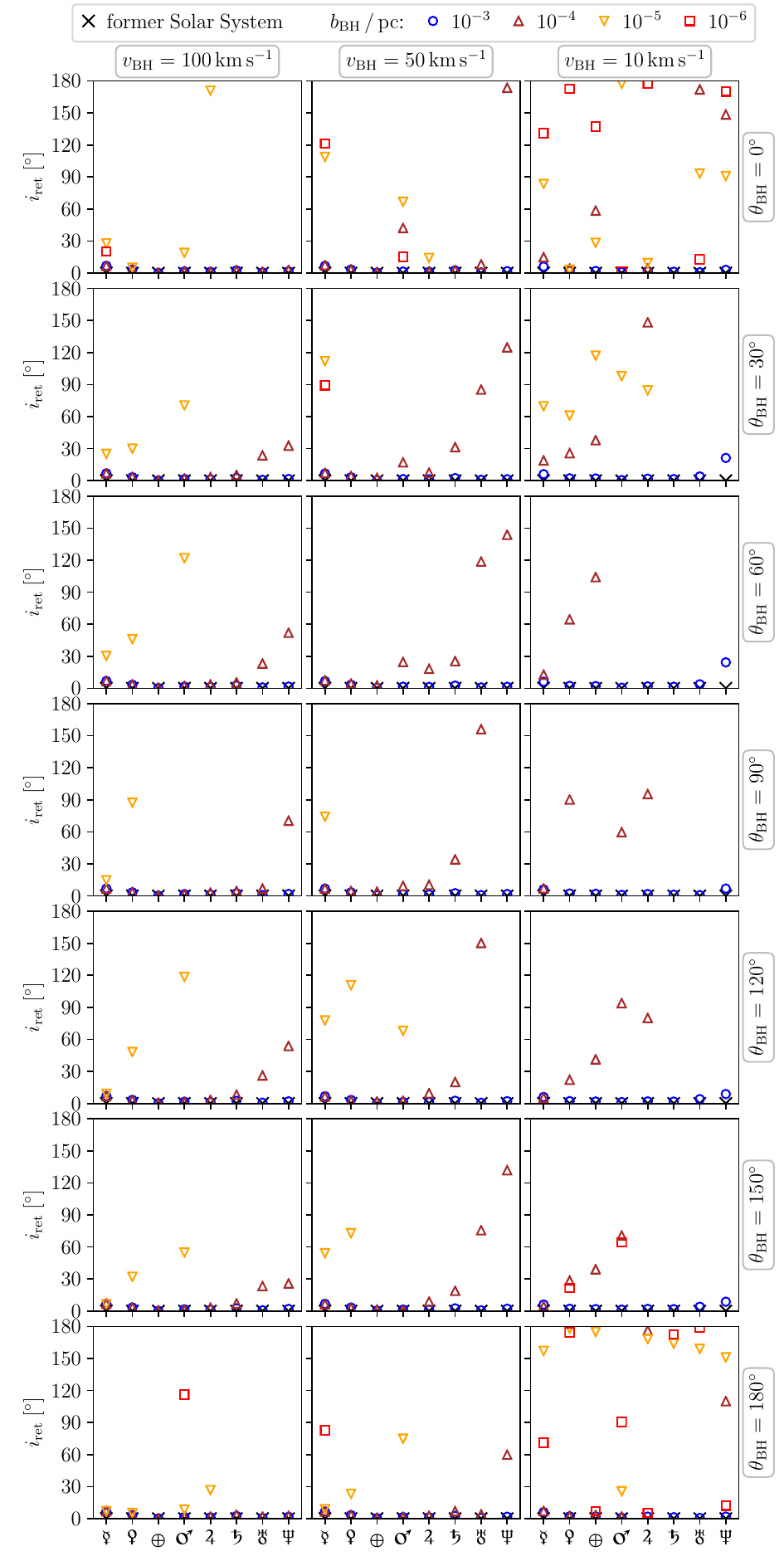}
	\caption{Distribution of inclinations of planets retained by the Sun after a~BH encounter. Each data point represents the mean value of the inclination of a~specific planet (Mercury~\Mercury\ to Neptune~\Neptune) if it was retained by the Sun. The data were compiled from all pre-evolved states of the SS for a~given set of initial conditions (impact parameter, $b_\BH$, encounter velocity, $v_\BH$, and the incidence angle, $\inc_\BH$) -- i.e. the panels do not show how many planets were retained in total (see Fig.~\ref{fig:bh_sun_ret}). The former SS is shown for reference.}
	\label{fig:bh_sun_inc}
\end{figure*}

\clearpage

\section{Additional figures: NS simulations}
\label{ap:ns}

\begin{figure*}[!htb]
	\centering
	\includegraphics[width=.8\linewidth]{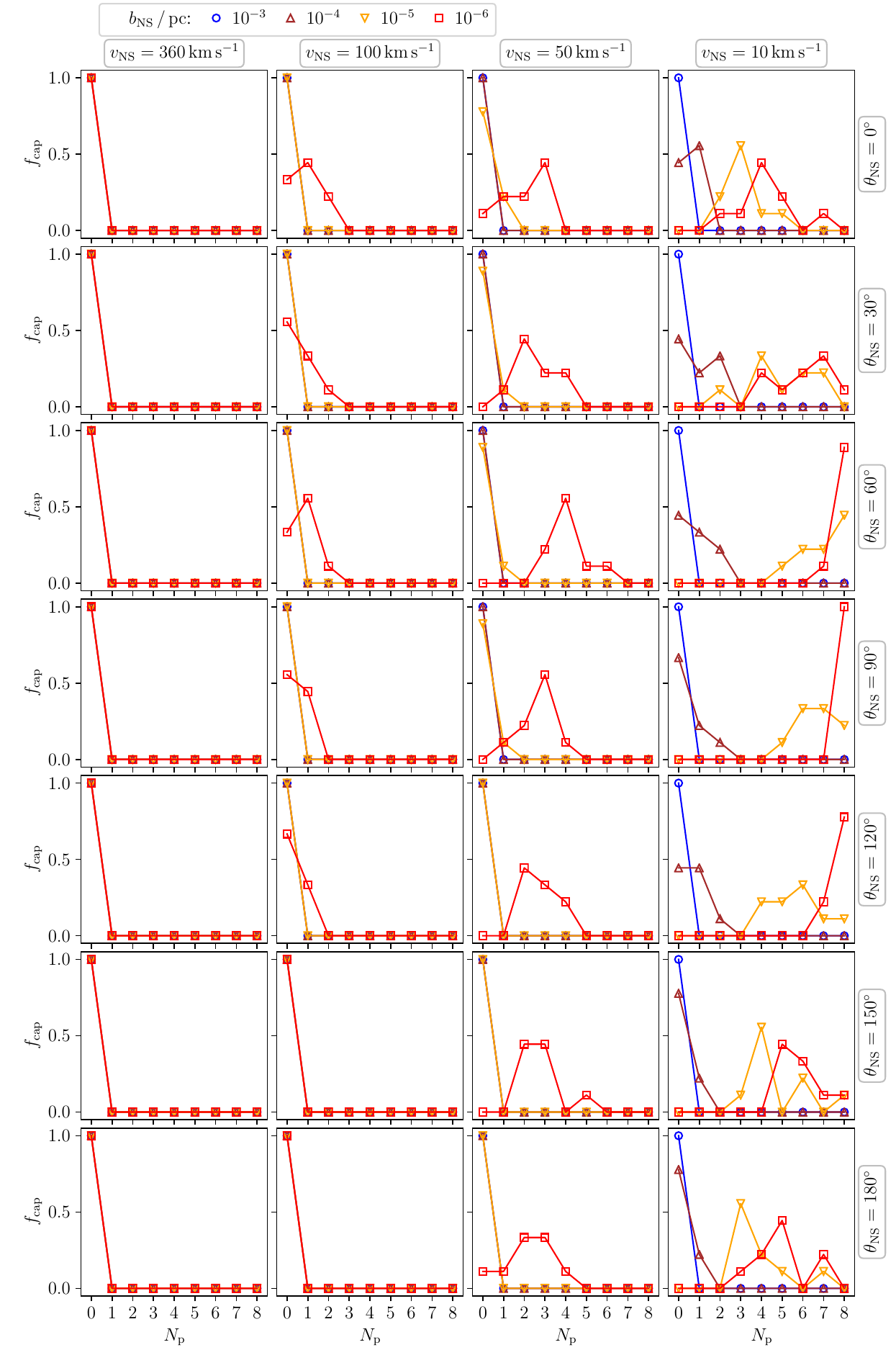}
	\caption{Capture of planets by a~NS. Frequency of capture, $f_\capture$, of a~given number of planets, $\Np$, by the NS for the specified set of initial conditions (impact parameter, $b_\NS$, encounter velocity, $v_\NS$, and the incidence angle, $\inc_\NS$).}
	\label{fig:ns_cap}
\end{figure*}

\begin{figure*}
	\centering
	\includegraphics[width=.8\linewidth]{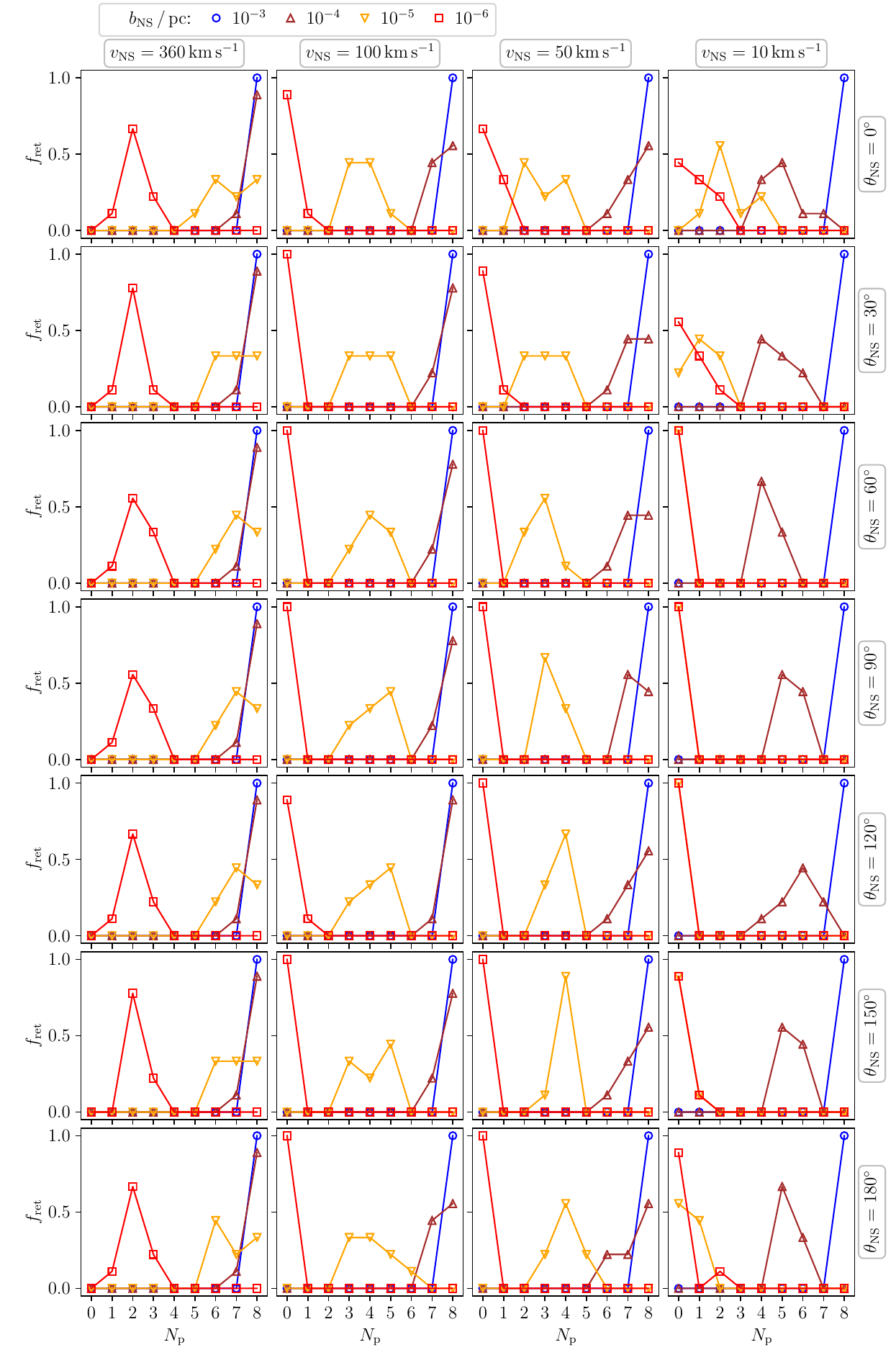}
	\caption{Planet retention after a~NS encounter. Frequency of retention, $f_\retain$, of a~given number of planets, $\Np$, by the Sun after the NS fly-by for the specified set of initial conditions (impact parameter, $b_\NS$, encounter velocity, $v_\NS$, and the incidence angle, $\inc_\NS$).}
	\label{fig:ns_sun_ret}
\end{figure*}

\begin{figure*}
	\centering
	\includegraphics[width=.8\linewidth]{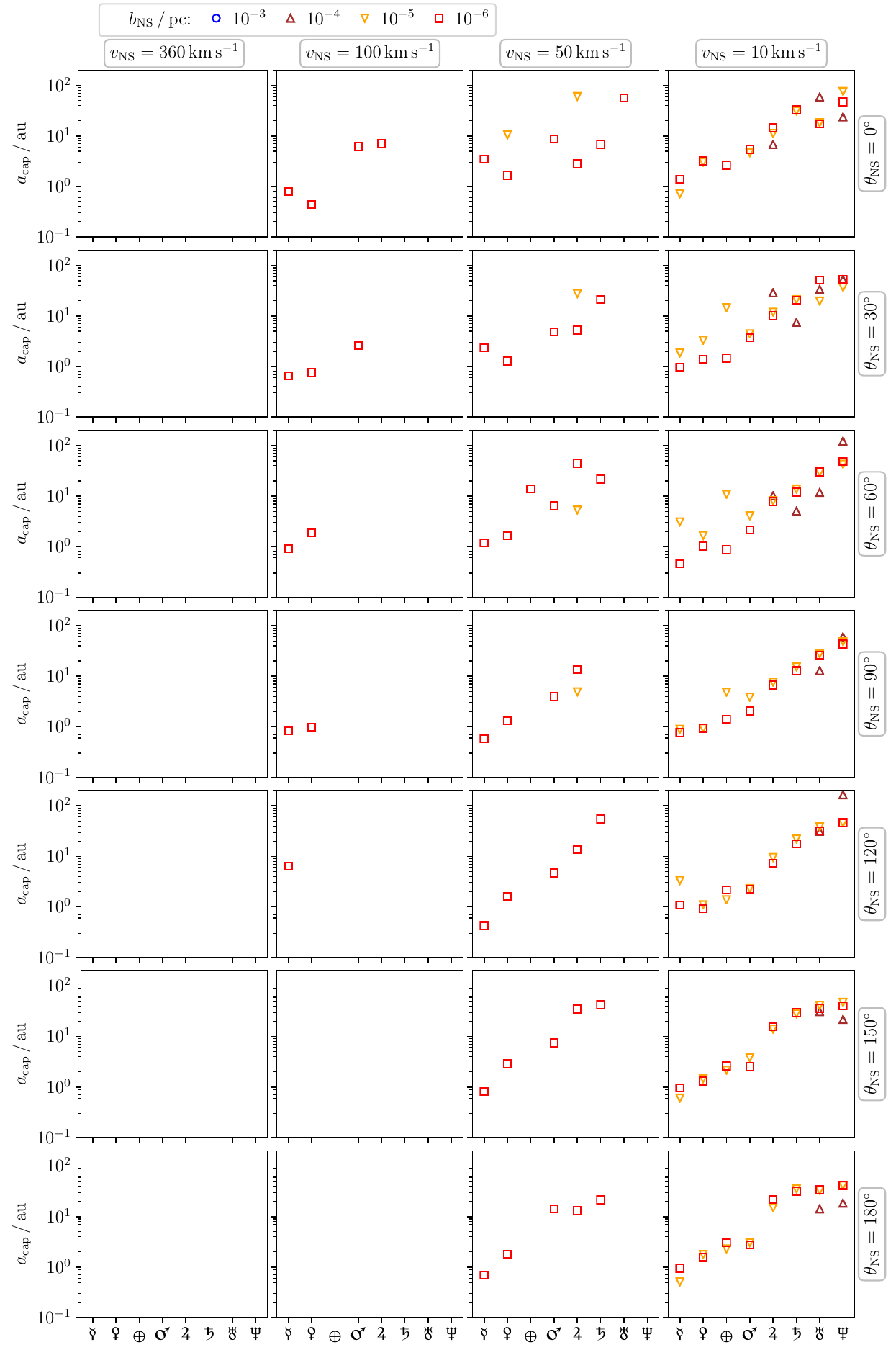}
	\caption{Distribution of semi-major axes of planets captured by a~NS. Each data point represents the mean value of the semi-major axis of a~specific planet (Mercury~\Mercury\ to Neptune~\Neptune) if it was captured by a~NS. The data were compiled from all pre-evolved states of the SS for a~given set of initial conditions (impact parameter, $b_\NS$, encounter velocity, $v_\NS$, and the incidence angle, $\inc_\NS$) -- i.e. the panels do not show how many planets were captured in total (see Fig.~\ref{fig:ns_cap} for reference). No planets were captured in the fastest scenario.}
	\label{fig:ns_semi}
\end{figure*}

\begin{figure*}
	\centering
	\includegraphics[width=.8\linewidth]{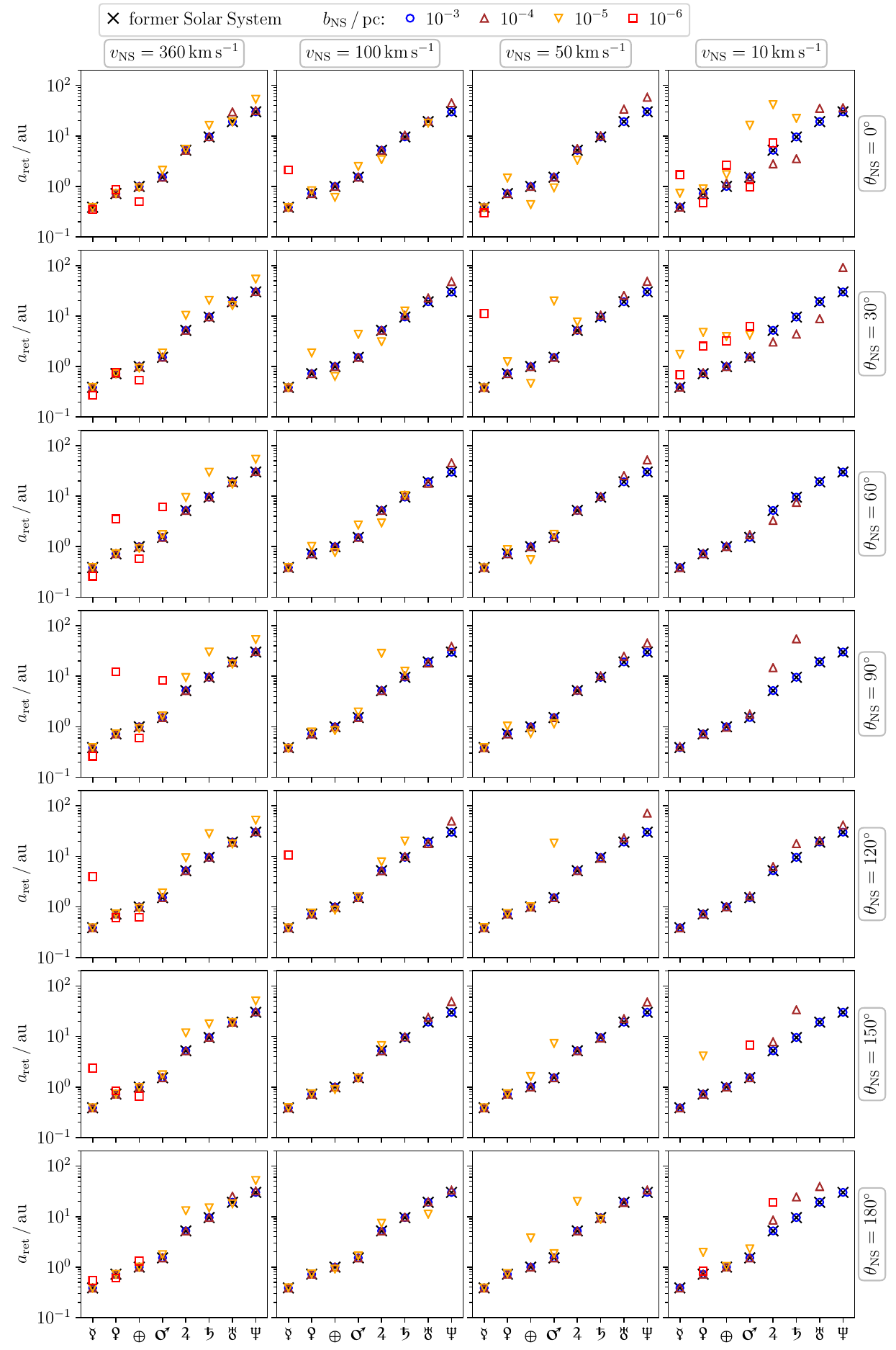}
	\caption{Distribution of semi-major axes of planets retained by the Sun after a~NS encounter. Each data point represents the mean value of the semi-major axis of a~specific planet (Mercury~\Mercury\ to Neptune~\Neptune) if it was retained by the Sun. The data were compiled from all pre-evolved states of the SS for a~given set of initial conditions (impact parameter, $b_\NS$, encounter velocity, $v_\NS$, and the incidence angle, $\inc_\NS$) -- i.e. the panels do not show how many planets were retained in total (see Fig.~\ref{fig:ns_sun_ret}). The former SS is shown for reference.}
	\label{fig:ns_sun_semi}
\end{figure*}

\begin{figure*}
	\centering
	\includegraphics[width=.8\linewidth]{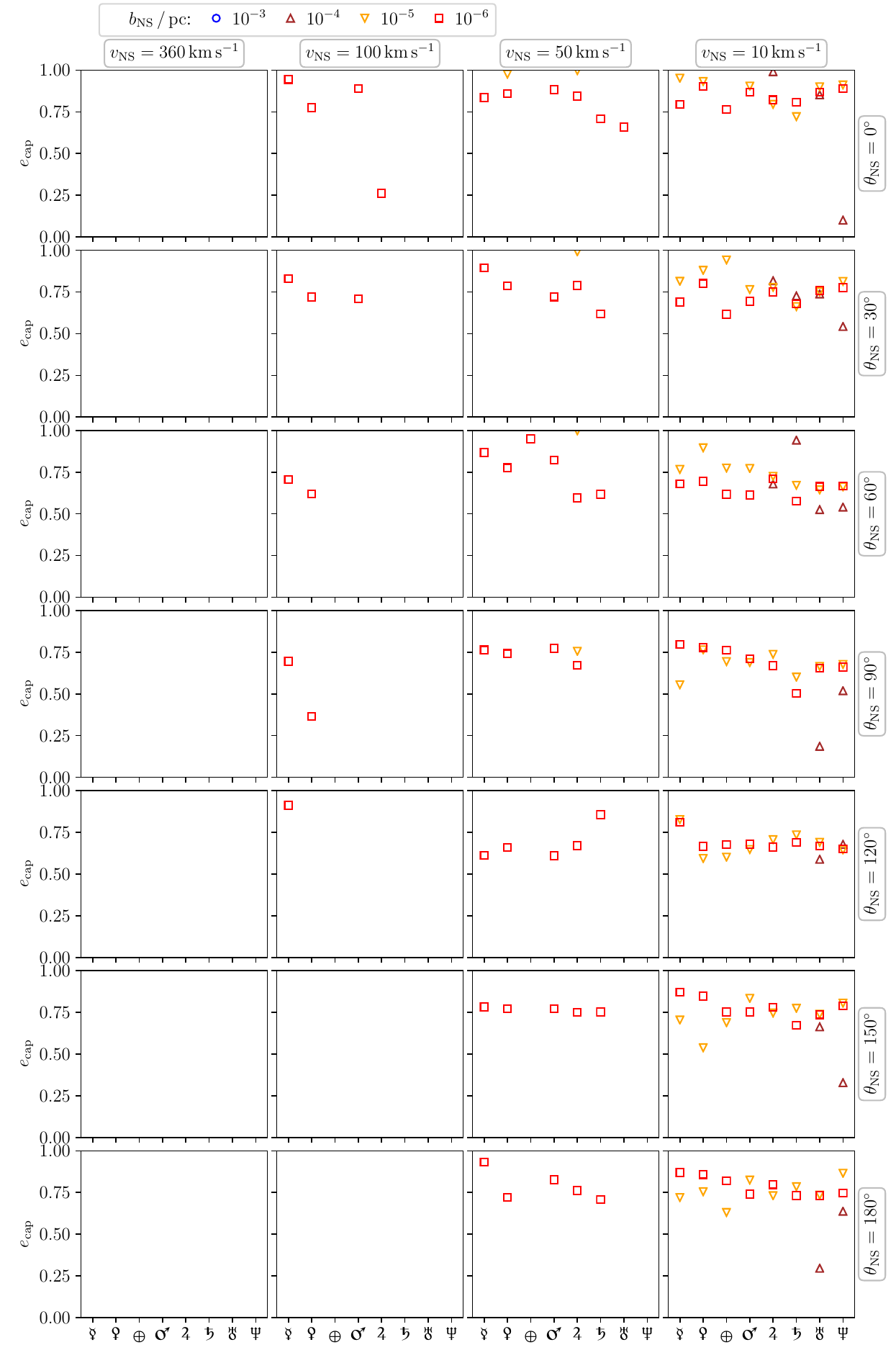}
	\caption{Distribution of eccentricities of planets captured by a~NS. Each data point represents the mean value of the eccentricity of a~specific planet (Mercury~\Mercury\ to Neptune~\Neptune) if it was captured by a~NS. The data were compiled from all pre-evolved states of the SS for a~given set of initial conditions (impact parameter, $b_\NS$, encounter velocity, $v_\NS$, and the incidence angle, $\inc_\NS$) -- i.e. the panels do not show how many planets were captured in total (see Fig.~\ref{fig:ns_cap} for reference).  No planets were captured in the fastest scenario.}
	\label{fig:ns_ecc}
\end{figure*}

\begin{figure*}
	\centering
	\includegraphics[width=.8\linewidth]{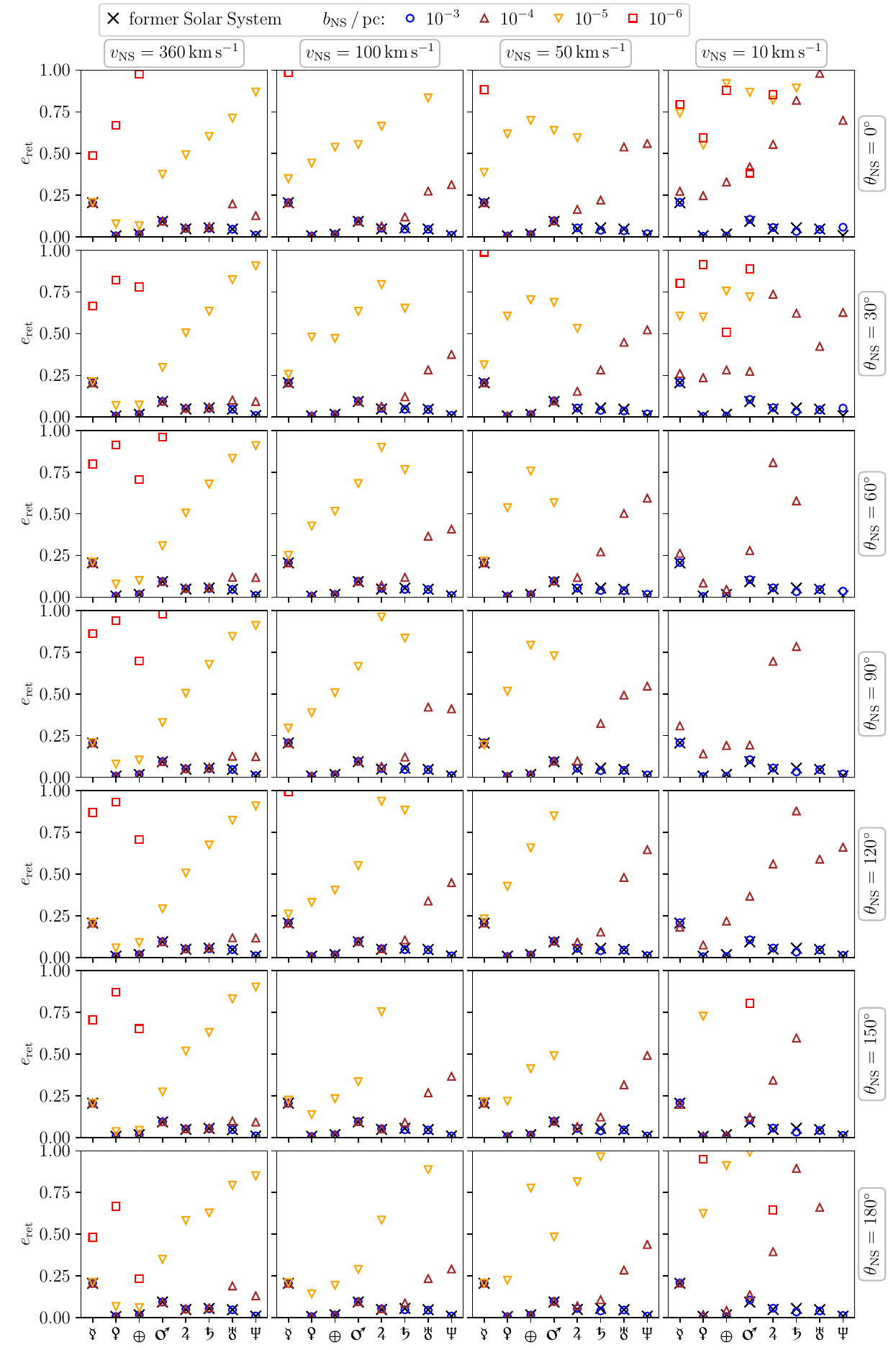}
	\caption{Distribution of eccentricities of planets retained by the Sun after a~NS encounter. Each data point represents the mean value of the eccentricity of a~specific planet (Mercury~\Mercury\ to Neptune~\Neptune) if it was retained by the Sun. The data were compiled from all pre-evolved states of the SS for a~given set of initial conditions (impact parameter, $b_\NS$, encounter velocity, $v_\NS$, and the incidence angle, $\inc_\NS$) -- i.e. the panels do not show how many planets were retained in total (see Fig.~\ref{fig:ns_sun_ret}). The former SS is shown for reference.}
	\label{fig:ns_sun_ecc}
\end{figure*}

\begin{figure*}
	\centering
	\includegraphics[width=.8\linewidth]{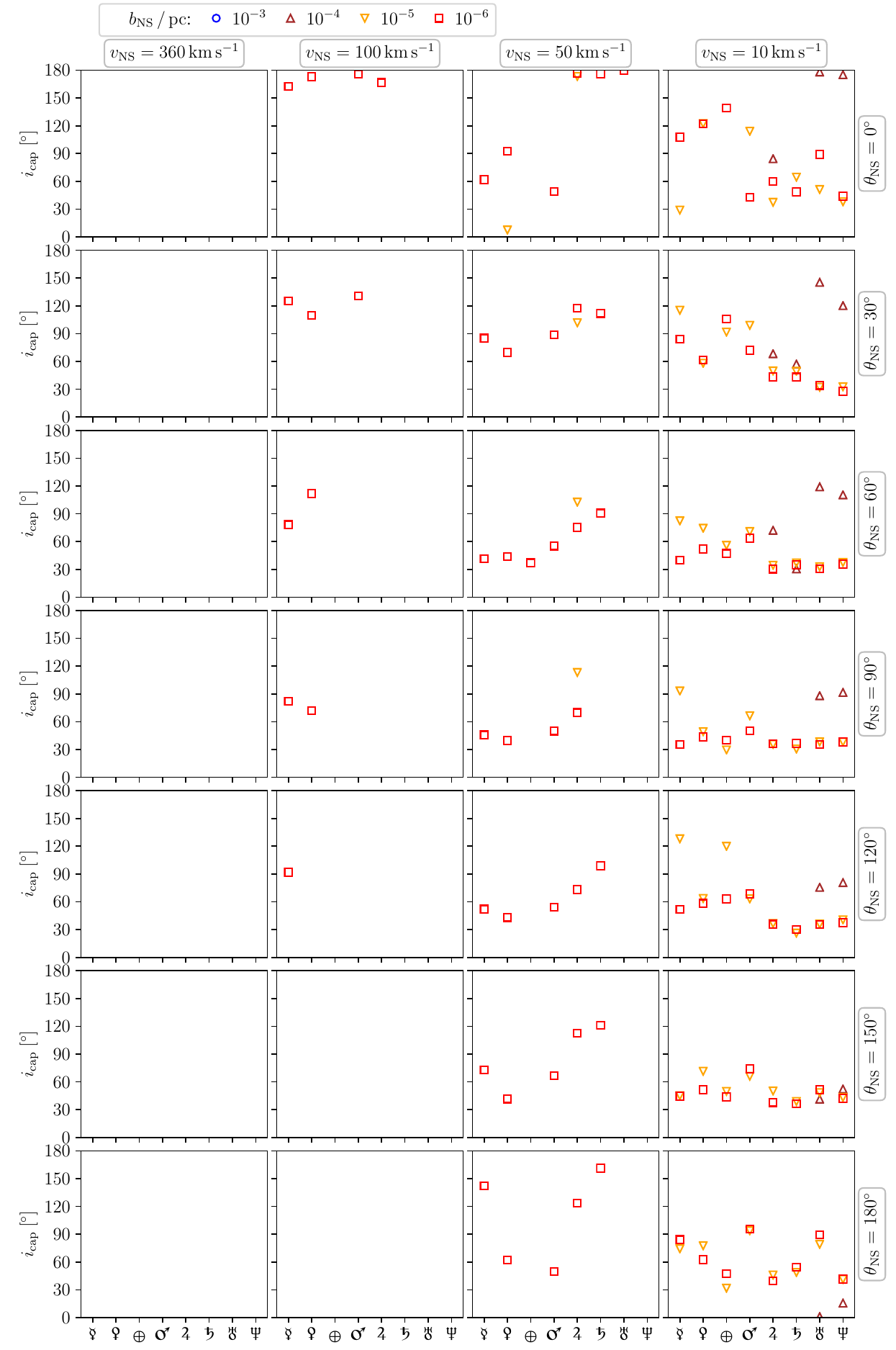}
	\caption{Distribution of inclinations of planets captured by a~NS. Each data point represents the mean value of the inclination of a~specific planet (Mercury~\Mercury\ to Neptune~\Neptune) if it was captured by a~NS. The data were compiled from all pre-evolved states of the SS for a~given set of initial conditions (impact parameter, $b_\NS$, encounter velocity, $v_\NS$, and the incidence angle, $\inc_\NS$) -- i.e. the panels do not show how many planets were captured in total (see Fig.~\ref{fig:ns_cap} for reference). No planets were captured in the fastest scenario.}
	\label{fig:ns_inc}
\end{figure*}

\begin{figure*}
	\centering
	\includegraphics[width=.8\linewidth]{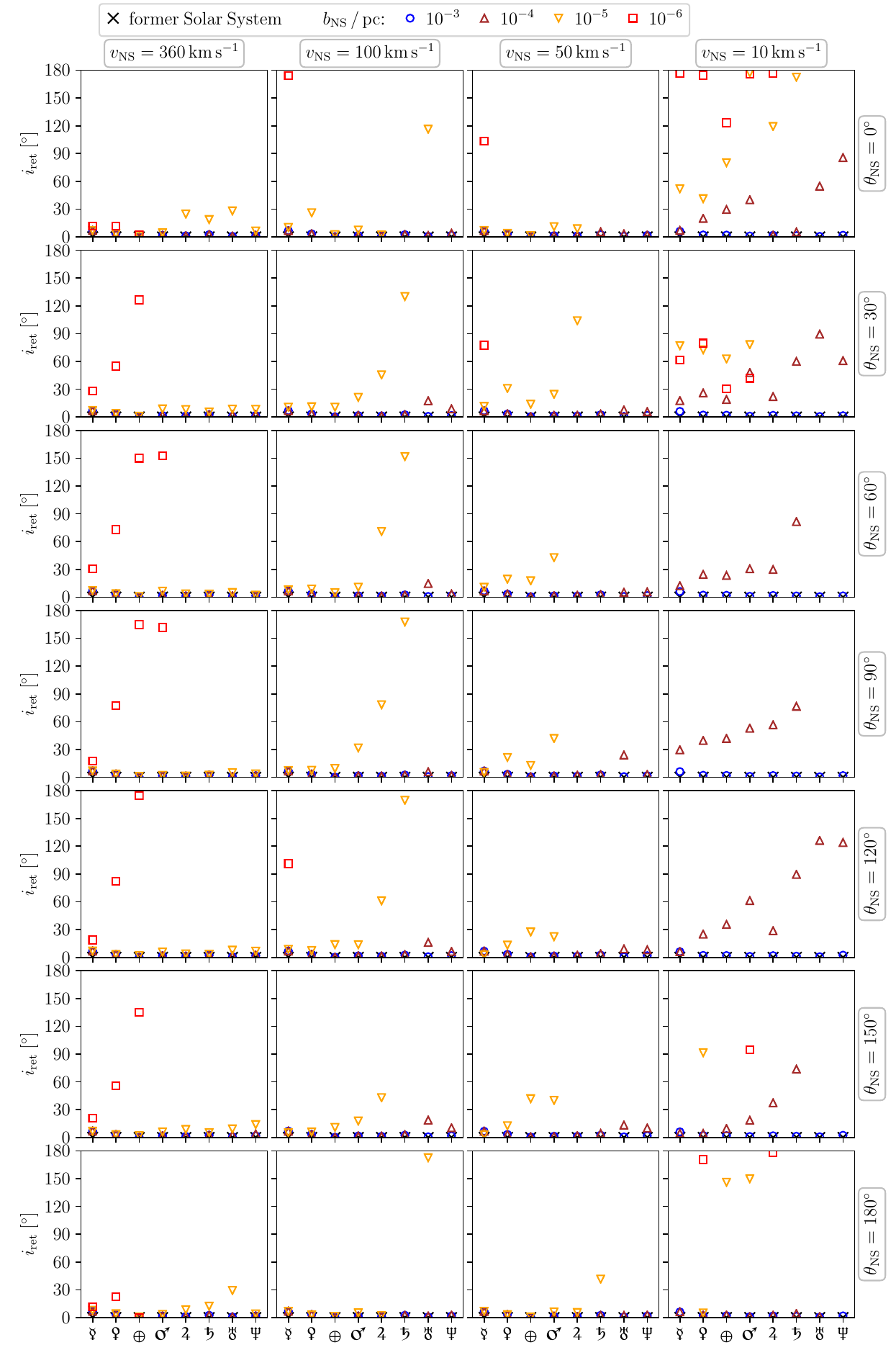}
	\caption{Distribution of inclinations of planets retained by the Sun after a~NS encounter. Each data point represents the mean value of the inclination of a~specific planet (Mercury~\Mercury\ to Neptune~\Neptune) if it was retained by the Sun. The data were compiled from all pre-evolved states of the SS for a~given set of initial conditions (impact parameter, $b_\NS$, encounter velocity, $v_\NS$, and the incidence angle, $\inc_\NS$) -- i.e. the panels do not show how many planets were retained in total (see Fig.~\ref{fig:ns_sun_ret}). The former SS is shown for reference.}
	\label{fig:ns_sun_inc}
\end{figure*}

\end{document}